\documentclass[aoas]{imsart}

\usepackage[english]{babel}
\usepackage{lmodern}
\usepackage[T1]{fontenc}

\usepackage{amsmath,amssymb}
\usepackage{graphicx}
\usepackage[colorlinks=true, allcolors=blue]{hyperref}
\newcommand{\appropto}{\mathrel{\vcenter{
  \offinterlineskip\halign{\hfil$##$\cr
    \propto\cr\noalign{\kern2pt}\sim\cr\noalign{\kern-2pt}}}}}

\newcommand{\E}{\mbox{$\mathbb{E}$}}

\newcommand{\cond}{\hspace{.01in}|\hspace{.01in}}

\newcommand{\p}{p}
\newcommand{\sig}{\mathsf{s}}
\newcommand{\back}{\mathsf{b}}
\newcommand{\sigCDF}{\mathsf{S}}
\newcommand{\backCDF}{\mathsf{B}}
\newcommand{\boldSigma}{\boldsymbol\Sigma}
\newcommand{\boldalpha}{\boldsymbol\alpha}
\newcommand{\boldLambda}{\boldsymbol\Lambda}

\let\hat\widehat
\let\tilde\widetilde

\usepackage{comment}
\usepackage{tikz}

\usepackage[authoryear,round]{natbib}
\begin{document}

\begin{frontmatter}
\title{COWs and their Hybrids: A Statistical View of Custom Orthogonal Weights}
\runtitle{COWs and their Hybrids}

\begin{aug}

\author{\fnms{Chad}~\snm{Schafer}},
\author{\fnms{Larry}~\snm{Wasserman}}
\and
\author{\fnms{Mikael}~\snm{Kuusela}}

\address{Department of Statistics \& Data Science and \\
STatistical Methods for the Physical Sciences (STAMPS) Research Center,\\ Carnegie Mellon University}

\end{aug}

\begin{abstract}
A recurring challenge in high energy physics is
inference of the signal component from a distribution
for which observations are assumed to be a mixture of
signal and background events.
A standard assumption is that there exists information
encoded
in a discriminant variable
that is effective at separating signal and background.
This can be used to assign a signal weight to each event, with these
weights used in subsequent analyses of
one or more control variables of interest.
The custom orthogonal weights (COWs) approach of
\cite{dembinski2022}, a generalization of the sPlot
approach of \cite{barlow1987} and \cite{pivk2005}, is tailored to address this
objective.
The problem, and this method, present interesting and
novel statistical issues. Here
we formalize the assumptions needed and the statistical properties, while also considering extensions and alternative approaches.
\end{abstract}

\begin{keyword}
\kwd{particle physics data analysis}
\kwd{signal separation}
\end{keyword}

\end{frontmatter}

\section{Introduction}

Detectors used in particle physics experiments collect
data on {\it events}, often resulting from
particle collisions within an accelerator, such as the Large Hadron Collider (LHC) at CERN. 
Some events comprise the
signal of interest, while others are contaminating background. For example, \cite{Aaij2025} reported on seminal results
from the LHCb Experiment \citep{LHCb} at the LHC in which key properties of the $\Xi_c$ baryon
were estimated,
including its
{\it spin-parity}, characterizing its total angular momentum and intrinsic symmetry under spatial inversion.
A central challenge to achieving these results is the
statistically rigorous separation of the $\Xi_c$ baryon (signal) events from
the background.

A common approach to achieving this separation (including that used in the
aforementioned study)
is the {\em sPlot} (or {\em sWeights}) method \citep{barlow1987,pivk2005}.
Here, the joint distribution of
a pair of random variables $(M,T)$
is modelled as 
\begin{equation}\label{eq::splot}
\p(m,t) = z g_1(m)h_1(t) + (1-z) g_2(m)h_2(t),
\:\:\: 0 \leq z \leq 1,
\end{equation}
where the first term represents the signal distribution
and the second represents the background. In the sPlot
formulation,
the functions $g_1, g_2, h_1, h_2$ are each densities,
but this could be relaxed.
$M$ is referred to as the {\em discriminant variable}
(often a mass measurement) and
$T$ is called the {\em control variable} (often an angular variable or some other kinematic property of the event).
In this model, $M$ and $T$ are independent in the signal and background.
That is, $M$ and $T$ are conditionally independent given $S$,
where $S\in\{1,2\}$ denotes signal or background.
The goal is to estimate 
the signal density for $T$, $\sig(t) =h_1(t)$.

In the absence of additional assumptions, the model (\ref{eq::splot})
is an example of a nonparametric mixture model
with conditional independence constraints.
Such models have been well-studied in the statistics literature. See, for example,
\cite{hall2005nonparametric, hall2003, chauveau2015, allman2009,zheng2020,
bonhomme2016}.
A surprising result, due to 
\cite{hall2003} and \cite{allman2009}, is that
the mixture (\ref{eq::splot}) is
nonparametrically identified if a third
conditionally independent variable is
added.
But with only two variables, and lacking further
assumptions,
the model (\ref{eq::splot}) is non-identified and we may find
other functions
$\tilde g_1, \tilde h_1, \tilde g_2, \tilde h_2$ and $\tilde z$ such that
$$
\p(m,t) = \tilde z \, \tilde g_1(m)\tilde h_1(t) + 
(1-\tilde z) \tilde g_2(m)\tilde h_2(t).
$$
Thus, recovering the component functions is impossible
even if the joint distribution $\p(m,t)$ were known.

The sPlot approach avoids this non-identifiability
by restricting the form of $g_1$ and $g_2$. In the
extreme case, these functions can be taken as fully
known, based on theoretical understanding of the 
underlying physics. A more realistic assumption,
often employed, is that a sample is available from
the discriminant variable, and assumptions are made
on the parametric forms of $g_1$ and $g_2$. The
parameters of these distributions (often called
{\it shape parameters} in particle physics) can hence
be estimated provided that the chosen parametric
forms result in an identified model. This sets up
the main heuristic of this
approach: Important information is assumed to be
known regarding the nature of the signal and
background in the discriminant variable $M$ and, from this
knowledge, events can be assigned {\it weights}
that quantify the likelihood the event is from
signal. These weights can then be used in analyses
applied to the control variable $T$ of interest.

But, this only works in the presence of the aforementioned
conditional independence assumptions.
\cite{dembinski2022}
introduced a more general model, called COWs
(custom orthogonal weights),
that relaxes this assumption as follows:
\begin{equation}\label{eq::cows1}
\p(m,t) = 
\underbrace{\sum_{k=1}^s z_{k}\, g_{k}(m)h_{k}(t)}_{\mathrm{signal}, \:\sig(m,t)} +
\underbrace{\sum_{k=s+1}^{b+s} z_{k}\, g_{k}(m)h_{k}(t)}_{\mathrm{background}, \:\back(m,t)}, \quad \sum_{k=1}^{b+s} z_k = 1.
\end{equation}
The main target of interest is the signal density for $T$, denoted
$\sig(t)$\footnote{Throughout this work, to ease notation, we will use argument(s) to a density to specify the corresponding random variables, i.e., we will write $\sig(t)$ instead of
$\sig_T(t)$, unless there is a chance of ambiguity.}, for which
$$
\sig(t) \propto \sum_{k=1}^s z_{k}h_{k}(t).
$$
This model reduces to sPlot when $s=b=1$.
But when $s,b > 1$, the conditional independence assumption
of sPlot is relaxed and a broader range of distributions can
be fit.
Critically,
the COWs method assumes that the densities ${\cal G}_1 =\{g_{1},\ldots, g_{s}\}$ and 
${\cal G}_2=\{g_{s+1},\ldots, g_{s+b}\}$ are specified, up to
at least a parametric form for each $g_{k}$, and that the resulting shape
parameters, along with the $z_{k}$,
are identified and can be estimated consistently.
%
Note that the marginals for the background and signal 
for $M$ can be
written as, respectively,
\begin{align}\label{eq::cows2}
\sig(m) &\propto \sum_{k=1}^s z_{k}\,g_{k}(m)\:\:\:\mbox{and}\:\:\:
\back(m) \propto \sum_{k=s+1}^{s+b} z_{k}\,g_{k}(m),
\end{align}
and, if $\sig(m)$ and
$\back(m)$ were taken as known, there would be
a constraint placed on the $g_{k}$'s.

When the $g_{k}$'s are given, 
$\sig(t)$ is identified.
But,  following the results of \cite{hall2003},
there are many choices of $g_{k}$ that satisfy
(\ref{eq::cows2})
and give the same joint $\p(m,t)$.
So, in fact, the COWs model is not, in general, identified.
However, that does not mean that
$\sig(t)$ is non-identified; indeed,
whether it is identified
remains an open question.
That is:
if we change the functions $g_{k}$ while preserving
$\p(m,t)$ and preserving the constraint (\ref{eq::cows2}),
does $\sig(t)$ remain the same?
If yes, then $\sig(t)$ is identified.
If no, then $\sig(t)$ is not identified.

The purpose of this paper is to review the COWs method,
to study its statistical properties as well as the
identification questions above and to consider
some extensions.
In particular, we consider an alternative way
to relax the conditional independence
assumption, namely, a mixture of
copulas. This approach avoids some of the identifiability issues.
In problems where there are more than two
variables, sPlot is fully nonparametrically
identifiable. Estimating such a model can be tricky
but COWs suggests a new method
for this problem
as we explain in Section
\ref{section::leverage}.

\section{Background}

In this section, we explain further the COWs approach.
Starting with the model given in Equation (\ref{eq::splot}),
the insight of the sPlot method was to note that a {\it weighting
function} $w_{\sig}$ with properties
\[
   \int w_{\sig}(m) g_1(m)\:dm = 1 \:\:\:\:\mbox{and}\:\:\:
   \int w_{\sig}(m) g_2(m)\:dm = 0
\]
would have the effect of isolating the signal component of
the control variable, i.e.,
\[
   \int w_{\sig}(m) \p(m,t) \:dm = z h_1(t).
\]
A weighting function for background, $w_{\back}$, could be similarly defined.
Hence, the recipe becomes as follows: (1) Either assume a fixed or
parametric form for $g_1$ and $g_2$; (2) If necessary, use a sample to estimate the shape parameters for $g_1$
and $g_2$; (3) Construct the weight function $w_{\sig}$ from (the estimates of) $g_1$ and $g_2$;
(4) Obtain weights for each observation based on $w_{\sig}$; and (5)
Perform the desired analysis on the sample in the control variable,
but incorporating the calculated weights.\footnote{There is no restriction
that the weights be nonnegative. This has led to the development of
methods specifically tailored to handle negative weights (e.g.,
\cite{Borisyak2019} and \cite{Andersen2022}). There is also not a unique $w_{\sig}$
function that satisfies the above conditions for fixed $g_1$ and $g_2$,
so sPlot chooses the function
which minimizes an estimate of ${\rm Var}(w_{\sig}(M))$.}

COWs takes this idea and extends it to a wider collection of
basis densities. Now, each function $g_{k}$ will have associated
with it a weight function $w_{k}$ such that
\[
   \int w_{j}(m) g_{k}(m) \:dm = \delta_{jk}.
\]
This implies that
\[
\int w_{k}(m) \p(m,t) \: dm = z_{k} h_{k}(t),
\]
and hence these weight functions extract the components
of the mixture corresponding to the control variable, and
\[
   \sig(t) \propto \sum_{k=1}^{s} z_{k} h_{k}(t)
   = \sum_{k=1}^{s} \int w_{k}(m) \p(m,t) \:dm
   = \int w_{\sig}(m) \p(m,t) \:dm,
\]
where
\[
   w_{\sig}(m) = \sum_{k=1}^{s} w_{k}(m)
\]
is the total weight across the signal weighting functions.

As demonstrated in \cite{dembinski2022}, the weighting functions
can be written as
\[
   w_{j}(m) = \sum_{k=1}^{b+s} \frac{a_{jk} g_k(m)}{I(m)}
\]
where $I(m)$ is an arbitrary non-zero function. The $a_{jk}$
are the entries of the matrix ${\bf A} = {\bf C}^{-1}$, where
the entries of ${\bf C}$ are the weighted inner products of the
$g_k$:
\begin{equation}
\label{eq::defc}
   c_{jk} = \int \frac{g_j(m)g_k(m)}{I(m)} \:dm.
\end{equation}
Setting $I(m)$ equal to the marginal distribution for $M$
is motivated by the finding that, with this choice,
\[
   \widehat z_k = \frac{1}{n}\sum_{i=1}^n w_k(M_i)
\]
is the maximum likelihood estimator of $z_k$, 
where the $M_i$'s are an observed sample in the discriminant variable,
of size $n$.
Indeed, it is possible to show that in this case
${\bf C}$ is the Fisher information matrix. Of course, one
would not know the marginal for $M$ in practice, and hence
$I(m)$ would be replaced with a suitable estimate.

Further approximation may be required in the estimation of
the $g_k$, which may only be specified up to the aforementioned
shape parameters. These estimated $g_k$ are plugged into 
Equation (\ref{eq::defc}) above, and the estimated inner products
are propagated forward.

\subsection{Flexibility of the COWs Model}
In \cite{dembinski2022}, it is claimed that
any joint density $\p(m,t)$
can be approximated with a COWs model
if a sufficient number of terms are included.
This is not quite true, as we now explain.
Let $g_1,g_2,\ldots$
be a basis for $L_2(\mathbb{R})$.
Then, assuming $\int\int \p^2(m,t) dm\, dt < \infty$,
we can write
$$
\p(m,t) = \sum_i \sum_j \beta_{ij}\, g_i(m)g_j(t)
$$
for some coefficients $\beta_{ij}$.
Hence,
$$
\p(m,t) = \sum_i g_i(m) h_i(t)
$$
where 
$h_i(t)=\sum_j \beta_{ij}g_j(t)$,
which is consistent with the form assumed by COWs.
This, however, ignores the mixture structure, i.e.,
the full model is
$$
\p(m,t) = z \sig(m,t) + (1-z) \back(m,t) 
$$
for signal and background distributions $\sig$ and $\back$.
If we expand $\sig$ and $\back$ in the basis, we have
$$
\sig(m,t) = \sum_i \sum_j \gamma_{ij}\,g_i(m) g_j(t),\ \ \
\back(m,t) = \sum_i \sum_j \beta_{ij}\,g_i(m) g_j(t) 
$$
for some coefficients $\gamma_{ij}$ and
$\beta_{ij}$.
Hence
\begin{align}
\p(m,t) &= \nonumber
z\sum_i\sum_j \gamma_{ij}\, g_i(m)g_j(t) +
(1-z)\sum_i\sum_j \beta_{ij}\, g_i(m)g_j(t)  
 \\
&= \label{eq::badcow}
z \sum_i g_i(m) \tilde h_i(t) + (1-z) \sum_i g_i(m) h_i(t),
\end{align}
where
\[
\tilde h_i(t)=\sum_j \gamma_{ij}g_j(t) \:\:\:\mbox{and}\:\:\:
h_i(t)=\sum_j \beta_{ij}g_j(t).
\]
Note, though, that the
$g_i$'s are the same functions in
the two sums in Equation (\ref{eq::badcow}).
The COWs method requires that the $g_i$ functions in
the two sums be linearly independent, otherwise there would not exist
weight functions that can extract the individual components $h_i$.

We conclude that while each of $\sig(m,t)$ and
$\back(m,t)$ can be written in COWs form,
the mixture requires strong restrictions
on them to maintain the conditions needed for the method.
This relates to questions about the identifiability of the model
given in Equation (\ref{eq::badcow}): It is not identified in
general, but is it identified subject to the
condition that both $\sig(m)$ and $\back(m)$, the signal and background distributions
in the discriminant variable, are known? (Or, more realistically, can be 
consistently estimated?)
In that case, one could impose the
additional constraints that
\[
   \sum_i \sum_j \gamma_{ij} g_i(m) \propto \sig(m) \:\:\:\mbox{and}\:\:\:
   \sum_i \sum_j \beta_{ij} g_i(m) \propto \back(m).
\]
Whether or not these constraints are sufficient for
identifiability is an open question.

\subsection{Synthetic Model}\label{section::Toymodel}

\begin{figure}[ht]
\centering
\includegraphics[width=5in]{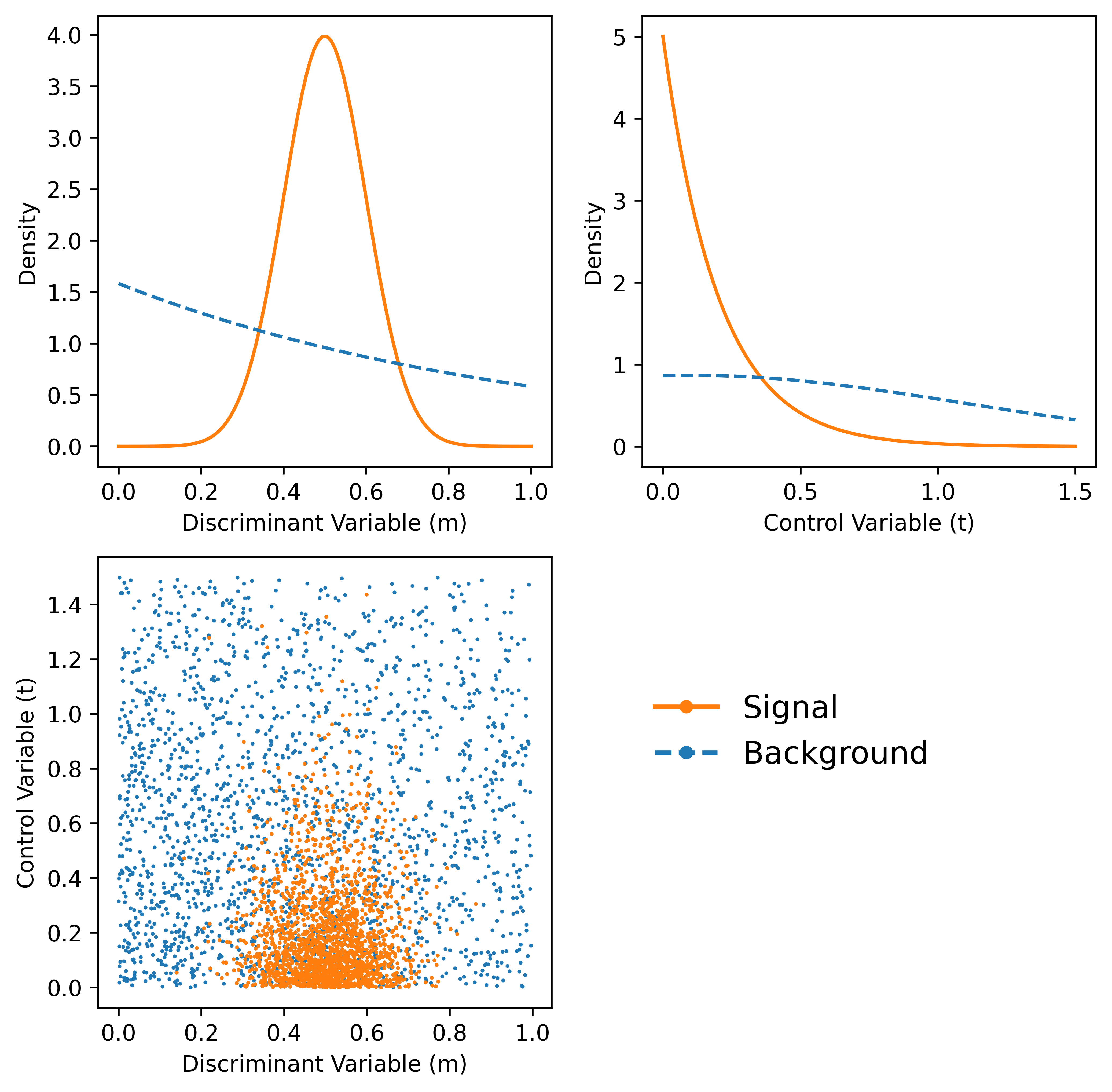}
\caption{The Synthetic model used throughout this work. The top two panels
show the signal and background distribution in both of the discriminant
and control variables. The bottom panel shows a sample drawn from this
distribution.}
\label{Syntheticmodel}
\end{figure}

As an illustrative
example, we will utilize a standard synthetic model used by
particle physicists.
Here, the signal in the discriminant variable has the
normal distribution with mean of 0.5 and standard deviation of 0.1, but the distribution
is truncated to the unit interval. The background in
the discriminant variable is the exponential distribution with mean 0.5, but
similarly truncated.
The signal in the control variable has the exponential distribution
with mean of 0.2, and the background in the control has the normal
distribution with mean of 0.1 and standard deviation of 1.0. 
These are both truncated
to $[0,1.5]$. 
The random variables $M$ and $T$ are independent given the signal/background label $S$.
Figure \ref{Syntheticmodel} shows these distributions, along with
a sample of simulated $(m,t)$ pairs. There are 2000 observations shown
from each of the signal and background.

\begin{figure}[ht]
\centering
\includegraphics[width=5.5in]{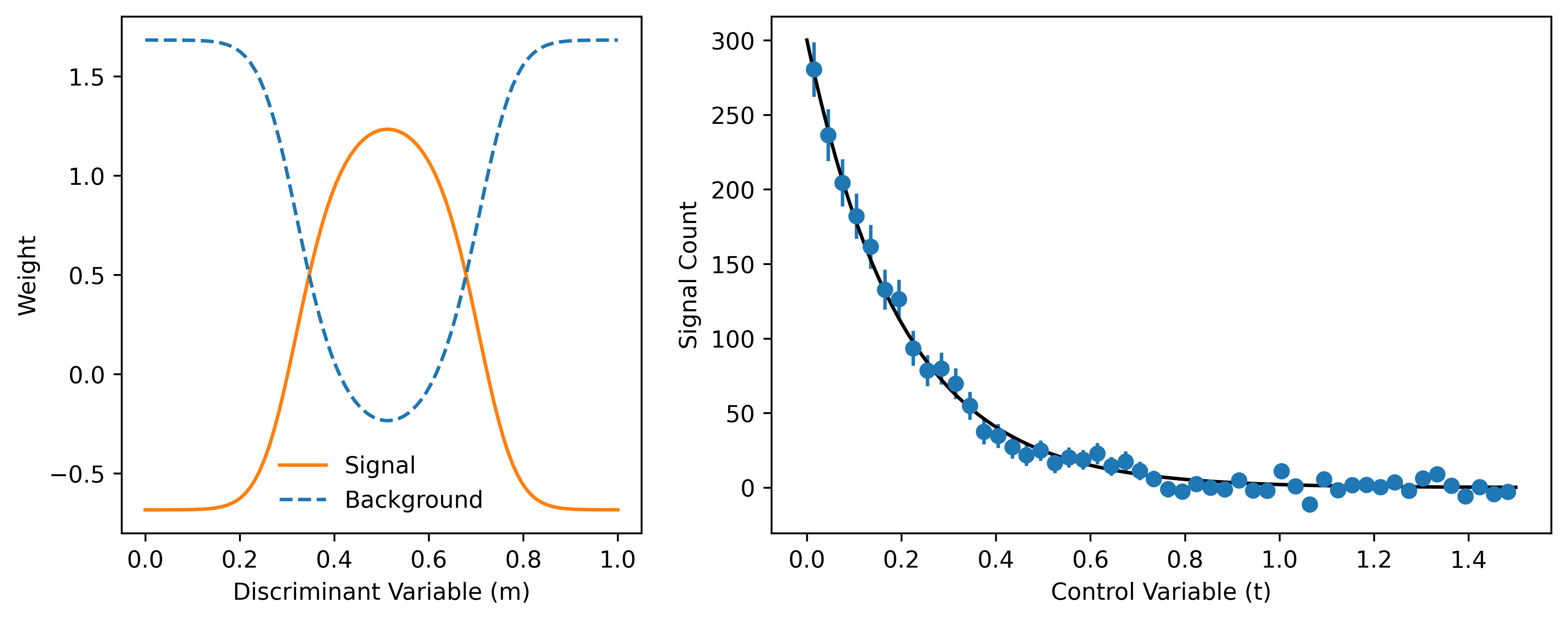}
\caption{Results from the application of sPlot/COWs to the simulated data set
shown in Figure \ref{Syntheticmodel}. The left panel shows the weight functions
$w_{\sig}$ and $w_{\back}$, corresponding to the signal and background, respectively.
The right panel shows the estimate of $nzh_1(t)$ that results when observations
are weighted using $w_{\sig}$. The solid line depicts the expected value of these
bin counts under the assumed model.}
\label{Syntheticresults}
\end{figure}

Figure \ref{Syntheticresults} shows the results when sPlot (COWs with $s=b=1$)
is applied to the data shown in Figure \ref{Syntheticmodel}. For this analysis,
it was assumed that the experimenter correctly specified that the signal
and background were normal and exponential, respectively, in $M$ but that the
shape parameters that determine these distributions were unknown. These
are hence estimated from the available data. This fitting is done via
maximum likelihood, where an additional estimated parameter is the
weight $z$ placed on the signal component in the mixture.

The weight functions $w_{\sig}$ and $w_{\back}$ are shown in the left
panel of Figure \ref{Syntheticresults}. Their forms are not surprising:
Observations with $m \approx 0.5$ are given the most weight
under the signal, and those with $m$ close to 0 or 1 are
given the most weight under the background.
There is a direct connection between these weights and the
estimated number of signal/background events. For example,
the total number of signal events is estimated as
\[
   n \widehat z = \sum_{i=1}^n w_{\sig}(m_i).
\]
The right panel of Figure \ref{Syntheticresults} uses this idea
to estimate the number of signal events in bins in the 
control variable, i.e., the estimated number of signal events
in the bin $B$ is
\[
   \sum_{i:\:t_i \in B} w_{\sig}(m_i).
\]
The solid line in this panel shows the value for $nz h_1(t)$ under the assumed model (scaled appropriately by the width of the bin used in the histogram).

\section{Related Approaches}\label{section::related}

This section explores some approaches to this challenge which
find their roots in the Statistics literature. Ultimately, each
makes different assumptions about the nature of the model and
the relationships between the variables.

\subsection{Mixture Weights}

It is natural to interpret the model used by COWs,
\begin{equation}
    \p(m,t) = \sum_{k=1}^{s+b} z_k g_k(m) h_k(t),
\label{eq::standardCOWs}
\end{equation}
as a description of observations that form a mixture of $s+b$ subpopulations,
i.e., $z_k$ equals the proportion of the observations that
arise from subpopulation $k$, and the joint distribution for $M$ and $T$ in
this subpopulation is $g_k(m)h_k(t)$. It follows that one could seek to estimate,
for each observation,
the probability that it arose from each of the
subpopulations, and then use the accumulated signal weights in
further analyses, in much the same manner as COWs. In fact, it is tempting to
interpret the weights resulting from COWs in a similar way, but clearly this is
not a valid interpretation given the potential for negative weights. Although,
in both cases, the sum of the weights is an estimate of the number
of observations drawn from subpopulation $k$, the COWs weights do not have a direct
probabilistic interpretation.

To formalize this idea, initially
assume that $z_1,z_2,\ldots,z_{s+b}$ and $g_1, g_2, \ldots, g_{s+b}$
are known.
The probability that observation $i$
was drawn from subpopulation $k$, conditional on $M_i$, is as follows:
\[
   w'_k(m_i) = P(\mbox{observation $i$ drawn from component $k \:|\: M_i = m_i$})
   = \frac{z_k g_k(m_i)}{\sum_j z_j g_j(m_i)}.
\]
If the $g_k$ were not known, but specified up to shape parameters, those parameters,
along with the $z_k$ could be estimated consistently provided linear independence
restrictions are satisfied by the $g_k$. The result would be approximations to
these weights.

Although this approach seems intuitive, it does make a fundamentally different
assumption than COWs. To explore this, start with
the simpler sPlot model given in Equation (\ref{eq::splot}).
Define
$$
w_1(m,t) = P(S=1 \cond M=m,T=t) = zg_1(m)h_1(t) /\p(m,t).
$$
Let $\delta$ be the Dirac delta function.
Then
$$
\E[ \delta(T-t) w_1(M,T)] =
z \int \frac{g_1(m) h_1(t)}{\p(m,t)} \p(m,t) \:dm =
z h_1(t),
$$
which recovers the marginal of interest.
But, we have access only to $w'_1(m)$ not $w_1(m,t)$ and we see that
\begin{align*}
\E[\delta(T-t)w'_1(M)] &=
\int w'_1(m)\p(m,t)dm \\
& =
\int \frac{w'_1(m)}{w_1(m,t)} w_1(m,t) \p(m,t) \:dm\\
&=
z \int \frac{w'_1(m)}{w_1(m,t)} g_1(m) h_1(t) \:dm\\
&=
z h_1(t) \int \frac{w'_1(m)}{w_1(m,t)} g_1(m)  \:dm
\end{align*}
which, in general, is not equal to $z h_1(t)$.
The mixture weighting method only is appropriate 
if $w'_1(m)=w_1(m,t)$, that is,
if $P(S=1 \cond M=m,T=t) =P(S=1 \cond M=m)$.
Thus, the mixture
weighting approach assumes
$S$ is independent of $T$ given $M$, while sPlot assumes that
$M$ and $T$ are independent given $S$.

\begin{figure}[ht]
\centering
\includegraphics[width=5.5in]{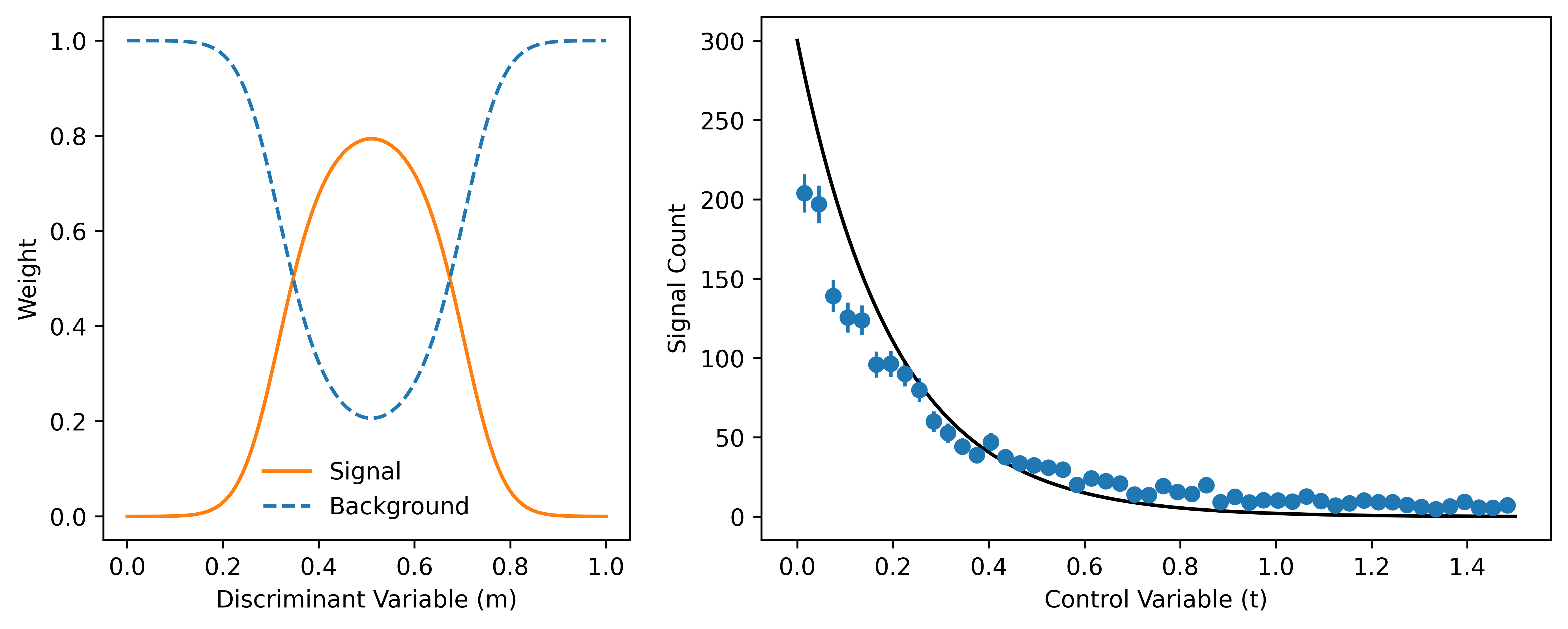}
\caption{Results from the application of mixture weighting approach
to the simulated data set
shown in Figure \ref{Syntheticmodel}. The left panel shows the weight functions
$w'_1$ and $w'_2$, corresponding to the signal and background, respectively.
The right panel shows the estimate of the signal distribution in the control
variable resulting from weighting
using $w'_1$. The solid line depicts the expected value of these
bin counts under the assumed model.}
\label{fig::toymodelinvweight}
\end{figure}

Here we will consider two examples which illustrate
the contrast between these assumptions.
First,
the simple model presented in Section \ref{section::Toymodel}
works well with the COWs approach but fails under mixture
weighting because the probability of an observation being signal
clearly depends on both the values of $M$ and $T$. See the
performance of the mixture weighting procedure in Figure
\ref{fig::toymodelinvweight}. The resulting bias is evident.

\begin{figure}[ht]
\centering
\includegraphics[width=4.5in]{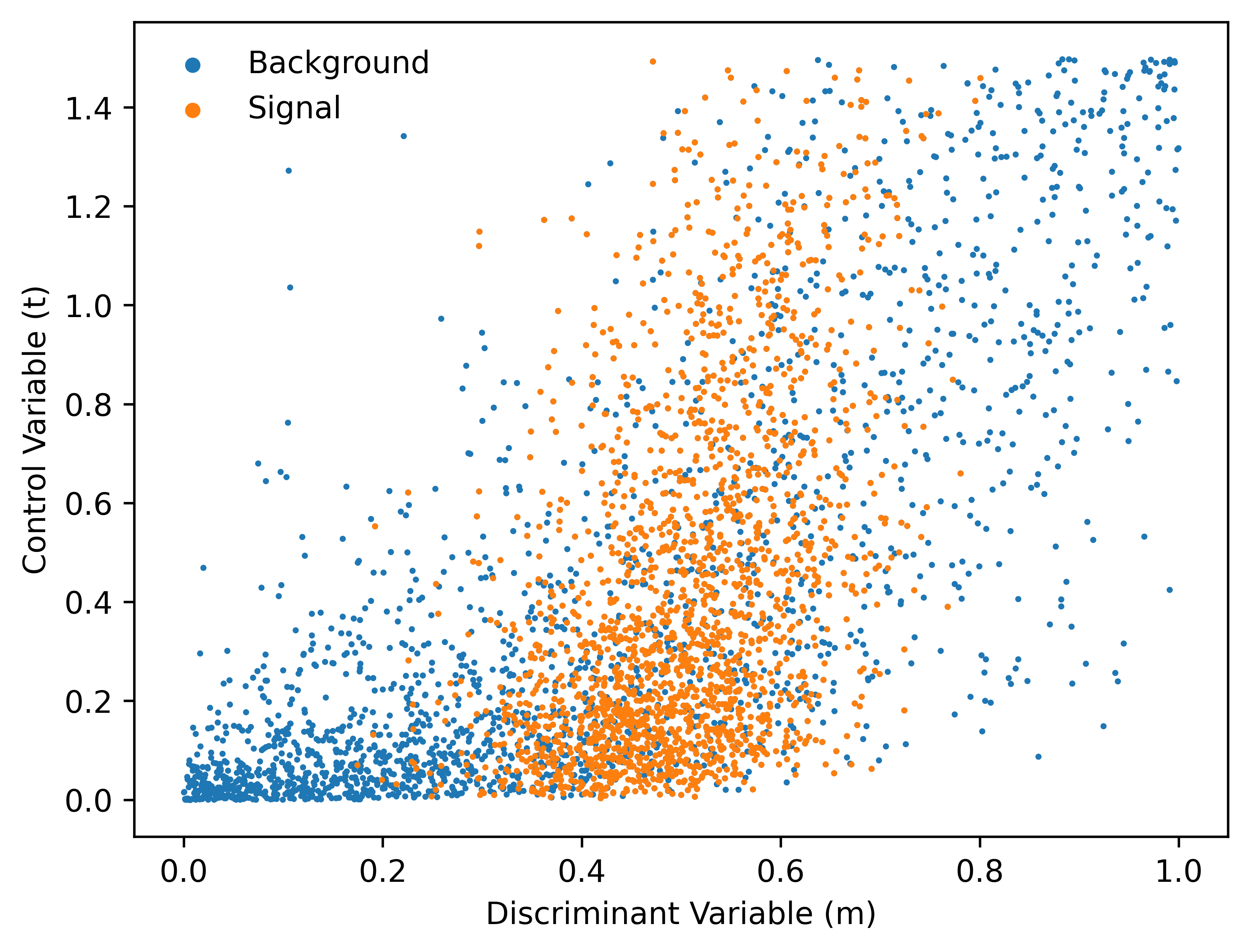}
\caption{Data simulated under the model with dependence between $M$ and
$T$. Note that within a slice of fixed value of $M$, the probability that
an observation is signal is independent of $T$.}
\label{fig::copulatoydata}
\end{figure}

For a second example, we will construct a case where
even though there
is a relationship between $M$ and $T$, the probability of an observation
being signal can be determined by only knowing the
value of $M$. Figure \ref{fig::copulatoydata}
shows an example of data simulated under such a model. This second example was constructed
so that the marginal distributions of $M$ and $T$ match those of the Synthetic model
used previously, but with dependence between $M$ and $T$ even when conditioning on $S$. 
This was accomplished
through the use of a {\it copula}, i.e., Sklar's Theorem \citep{sklar1959} states that
the joint distribution between $M$ and $T$ can be
decomposed as a product of the marginals and the copula density $c$:
\[
   \p(m,t) = \p(m) \p(t) c(F(m),F(t)).
\]
Here, a Gaussian copula is utilized with correlation parameter set to 0.7.
After the full sample of $(m,t)$ pairs were generated, then
each pair is randomly assigned to be either signal or background by using only the assumed signal and
background distributions for $M$. In other words,
in a vertical slice in Figure \ref{fig::copulatoydata}, the proportion of
observations which are signal will vary with $m$,
but the probability of it being signal will not
depend on $t$.

\begin{figure}[ht]
\centering
\includegraphics[width=5in]{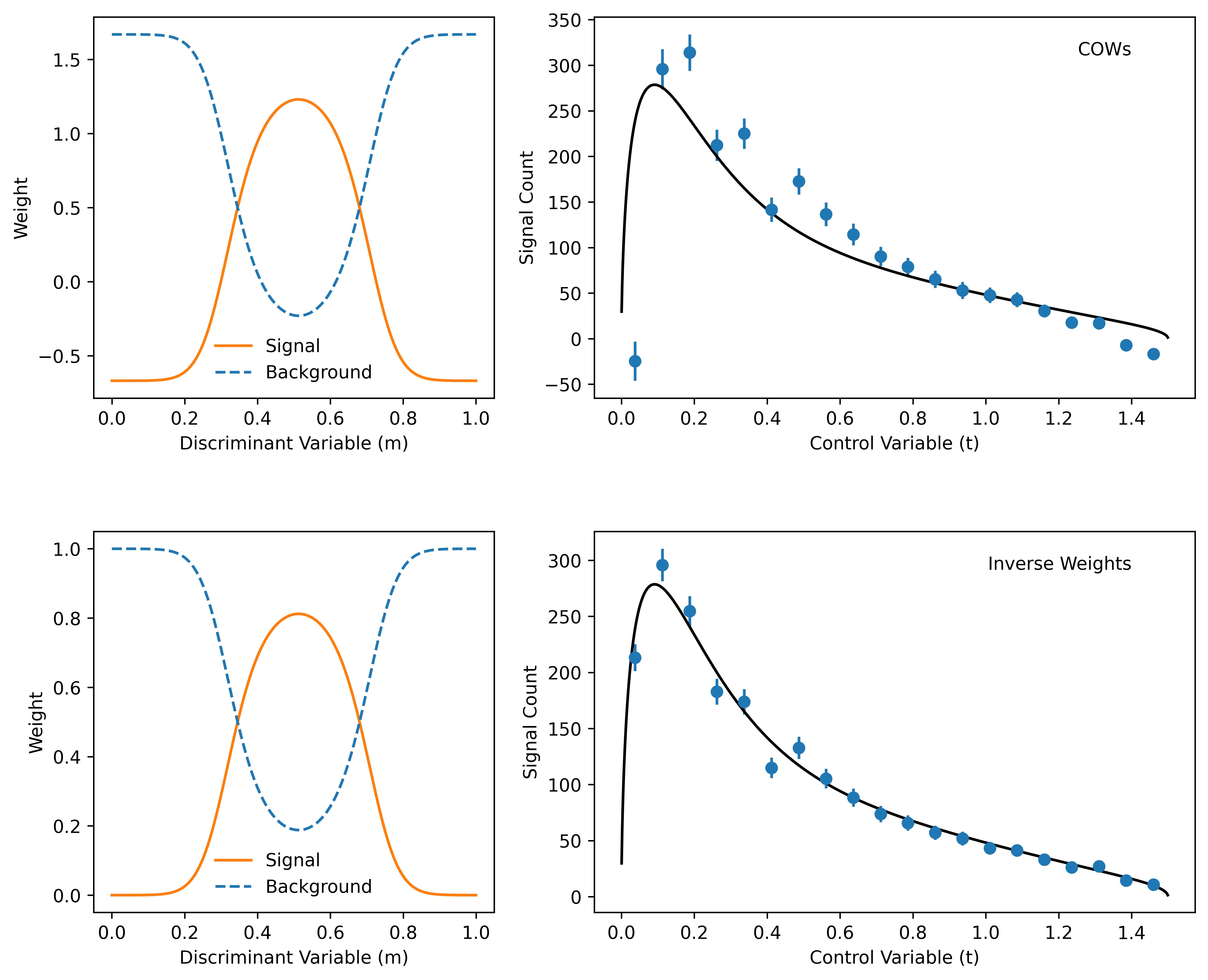}
\caption{Results when using both COWs and mixture weighting on the
data shown in Figure \ref{fig::copulatoydata}.}
\label{fig::copulatoyresults}
\end{figure}

Results using both COWs and mixture weighting are shown in
Figure \ref{fig::copulatoyresults}.
The solid line is the target of inference, as before, $z \p(t \cond S=1)$.
Under the present assumptions, it can be shown that
\begin{eqnarray*}
    \p(t \cond S=1) & = & \int \p(t \cond S=1, M=m) \p(m \cond S=1) \:dm \\
    & = & \int \p(t \cond M = m) \p(m \cond S = 1) \:dm \\
    & = & \int \left(\p(m,t)/\p(m)\right) \p(m \cond S=1) \:dm \\
    & = & \p(t) \int c(F(m), F(t)) \:\p(m \cond S=1) \:dm.
\end{eqnarray*}

It is reasonable to ask, even if
$P(S=1|M=m,T=t) =P(S=1|M=m)$ does not hold exactly,
might we expect it to hold approximately
in these applications? In order for this to be valid,
it must be the case that
$M$ has much more information about the mixture
structure than $T$ does.
In these cases, we expect
$w_1(M,T) \approx w_1'(M)$
in which case
$\E[I_t(T)w_1'(M)] \approx z h_1(t)$.
In Section \ref{section::copula} we will build on this idea of using
copulas to incorporate dependence between $M$ and $T$.

\subsection{Varying Coefficient Models}\label{sec::varycoef}

\begin{figure}[ht]
\centering
\includegraphics[width=4.5in]{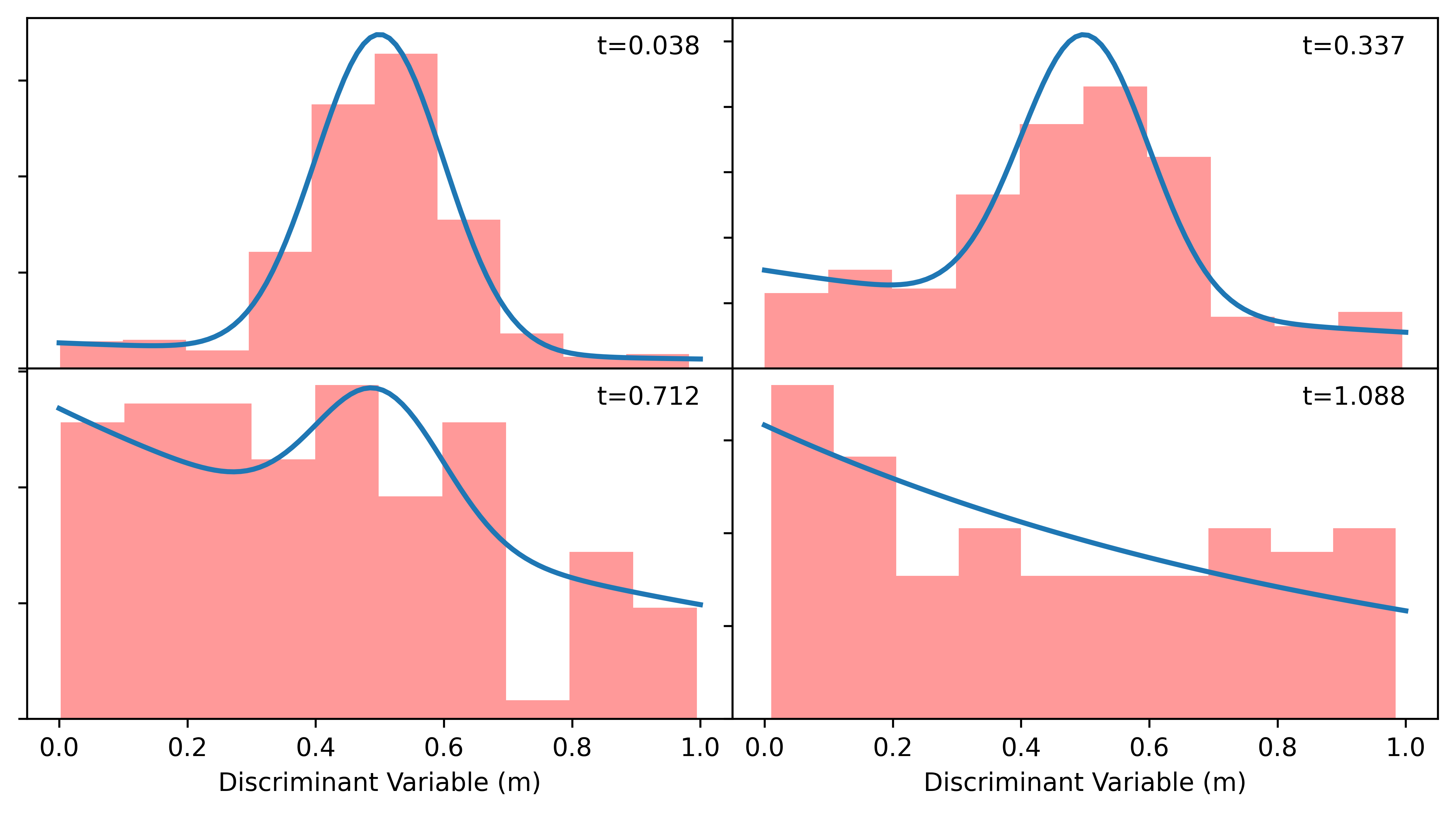}
\caption{Illustration of how the model fit evolves across
different values of $t$.}
\label{fig::mixhists}
\end{figure}

The model given in Equation (\ref{eq::standardCOWs}) can be
rewritten in terms of the conditional distribution
of $m$ given $t$, as follows:
\[
   \p(m \cond t) = \sum_{k=1}^{s+b} g_k(m) \beta_k(t),
\]
where
\[ 
   \beta_k(t) = \frac{z_k h_k(t)}{\p(t)}.
\]
This is a {\it varying coefficient model} \citep{Hastie1993} 
adapted for conditional density estimation.
Writing it in this way highlights the fact that each
slice in $t$ is being modelled as a mixture of the same
$s+b$ basis densities (the $g_k$). The function $\beta_k$
provides the coefficients (weights) that vary as a function of $t$.
Figure \ref{fig::mixhists} reconsiders the Synthetic model
presented above, showing how the model fit changes with the
value of $t$. In this case, there are only two basis functions
utilized, namely the case where the user has correctly specified
the signal and background components of the mixture. Similar
results would be obtained in the case where the signal and
background components are specified correctly, but with
shape parameters estimated from the available data.

Of course, when the model is viewed in this manner, it seems more
realistic that the $g_k$ could be chosen to enable a more flexible
model, and impose less restrictive assumptions on its form.
\citet{IzbickiLee2017} considered such a model for the
purpose of conditional density estimation, under the assumption
that the basis functions are chosen to be orthogonal, allowing
for the fitting of a wide range of distribution shapes.
Orthogonality leads directly
to a way to estimate the $\beta_k$
by projecting the observed data onto the $g_k$. However, for such an approach
to be practical in this setting, one needs to have a way of
separating the basis functions into those that yield the signal density and those that yield the background.
Realistically, such a justification must arise from consideration
of the physics underlying the situation.


\subsection{Nonnegative Matrix Factorization} \label{sec:NMF}

An alternative way to approach the signal and background separation
problem that motivated COWs is one of {\it blind source
separation}, a class of methods designed specifically
for separating signals in an unsupervised manner with minimal
assumptions. (See \cite{comon2010handbook} for a review.)
In particular, this situation is well-suited to using
{\it nonnegative matrix factorization (NMF)} \citep{Lee1999}.
The objective of NMF is to
approximate an $m$ by $n$ nonnegative matrix ${\bf P}$ as the product of two nonnegative matrices,
${\bf P} \approx {\bf GH}$,
where ${\bf G}$ and ${\bf H}$ are also each nonnegative and
of rank $r$, which is often chosen much smaller than either $m$ or $n$.
A standard approach to the estimation is to use numerical optimization to find ${\bf G}$
and ${\bf H}$
that minimize the Frobenius norm
$\| {\bf P} - \nobreak {\bf GH} \|_F^2$, although other norms can be used.

\begin{figure}[ht]
\centering
\includegraphics[width=5.5in]{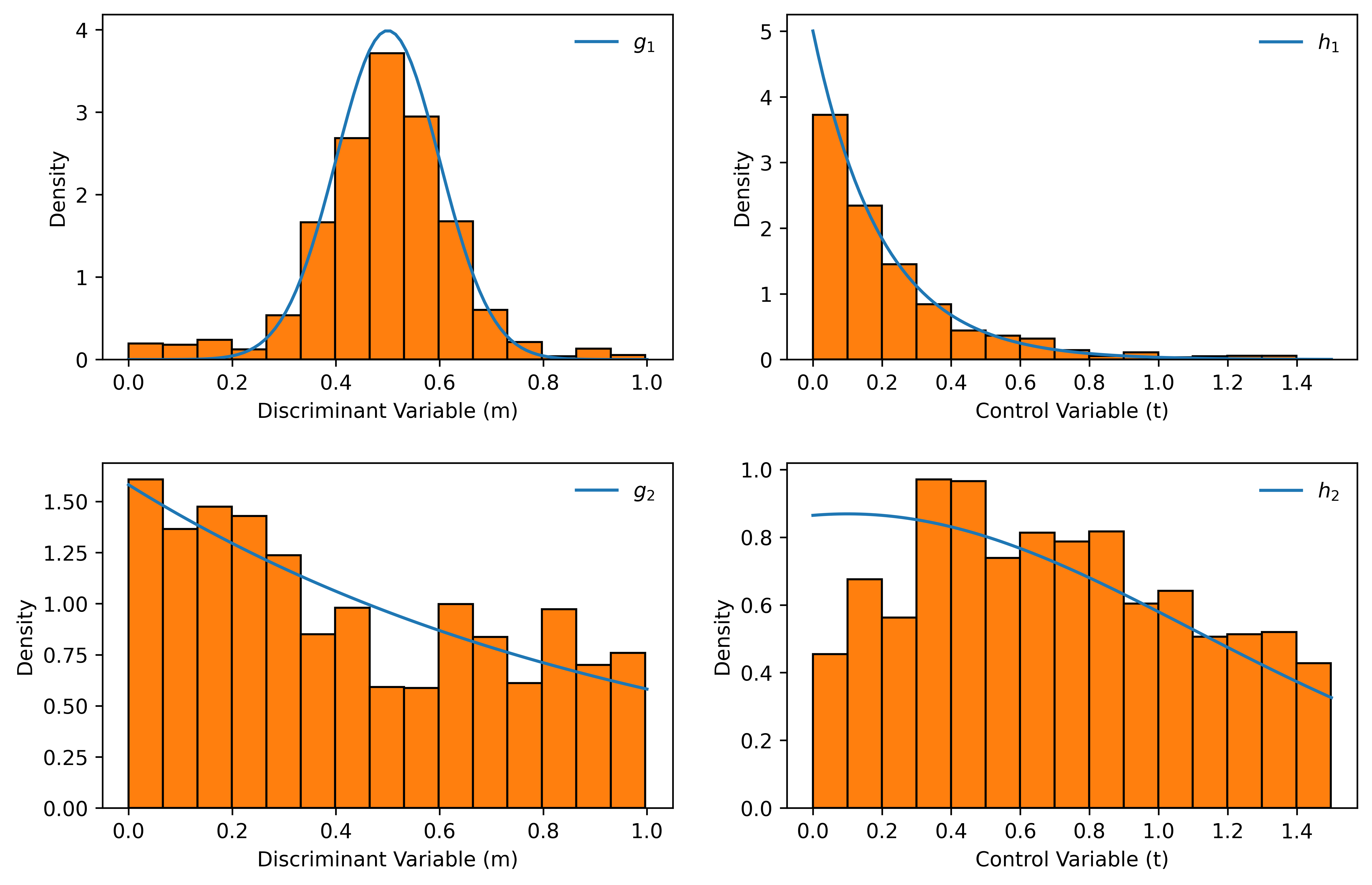}
\caption{The results of nonnegative matrix factorization when
applied to the Synthetic data. The first row compares
${\bf g}_1$ with $g_1$ and ${\bf h}_1$ with $h_1$. 
The second row compares
${\bf g}_2$ with $g_2$ and ${\bf h}_2$ with $h_2$.}
\label{fig::nnmfcomps}
\end{figure}

In the problem at hand, the matrix ${\bf P}$ would be constructed
as a two-dimensional histogram in $M$ and $T$, using the observed
data.
Note that
\[
   {\bf P} \approx {\bf G}{\bf H} = \sum_{k=1}^r {\bf g}_k {\bf h}_k,
\]
where ${\bf g}_k$ and ${\bf h}_k$ are the $k^{\mathrm{th}}$ column of
${\bf G}$ and row of ${\bf H}$, respectively. Then the similarity
between NMF and the COWs model of Equation (\ref{eq::standardCOWs})
begins to become evident. The joint distribution of $M$ and $T$,
now discretized, is being decomposed into a sum of outer products
of $({\bf g}_k, {\bf h}_k)$ pairs. Each pair can be viewed as
histograms using the same binning
as was used to create ${\bf P}$. 
In the case of our Synthetic example, the observed data were converted
into a histogram with $15 \times 15$ bins.
Figure \ref{fig::nnmfcomps} compares the vectors ${\bf g}_k$ and ${\bf h}_k$
with the actual assumed distributions $g_k$ and $h_k$. (Each vector was
normalized to yield a well-defined histogram.)
The ability of NMF to extract the distributions of interest,
i.e., those corresponding the signal, is impressive.


Nevertheless, practical concerns regarding the use of NMF
arise. First, the solution is not unique, in general.
An equivalent approximation can be derived
by finding an invertible matrix ${\bf M}$ such that
$\tilde{\bf G} = {\bf G} {\bf M} \geq 0$ and
$\tilde{\bf H} = {\bf M}^{-1}{\bf H}  \geq 0$.\footnote{Conditions under
which NMF yields a unique solution, not met
in this situation, can be found in \cite{Laurberg2008}.}
The superior performance of NMF in this case is enabled by the fact that
the primary source of variation in the matrix ${\bf P}$ can be
attributed to the signal. (In other words, the signal distributions
are much further from uniform than is the background.) The extent to
which this is true in real situations may largely dictate the value
of NMF in this setting.
This relates to a broad concern of how one would determine
which of the components reported
by NMF are associated with signal, and which are associated
with background. Without further assumptions, the method cannot help
with this question. Finally, the NMF procedure must be enhanced with
an approach to obtaining error bars on the histogram. This could be
accomplished via the bootstrap.

NMF is a well-studied procedure and there are extensions that may be
useful for this application. In particular, there has been work on
Bayesian approaches
in which one could place a prior on ${\bf G}$ and ${\bf H}$
\citep{Nakajima2011,Schmidt2009} or
impose sparsity, smoothness, or other constraints on the
solution \citep{Pauca2006,Maffettone_2021}.
In particular, constraints of the form
\[
   \sum_{k=1}^s {\bf g}_k = {\bf s}
   \:\:\:\:\mbox{and}\:\:
   \sum_{k=s+1}^{s+b} {\bf g}_k = {\bf b}
\]
could be of interest.
Any of these approaches could allow for the incorporation of
physical theory in order to generate a unique solution.

\section{COW(s) Tools}

This section proposes a few statistical enhancements to COWs
to further its applicability.

\subsection{Goodness-of-Fit Test}

A natural question is the suitability of the chosen set
of $g$ functions for modeling the distribution. For this
purpose, it is useful to reconsider the varying
coefficient model formulation presented in Section \ref{sec::varycoef}. Specifically, the model assumes that
for each slice in $T$, the conditional distribution
$\p(m \cond t)$ can be well-approximated as a mixture
of the chosen, or estimated, functions $g_1,g_2,\ldots, g_{s+b}$. The comparisons made in Figure \ref{fig::mixhists}
illustrates this construction for four different values of
$T$.

\begin{figure}[ht]
\centering
\includegraphics[width=5.5in]{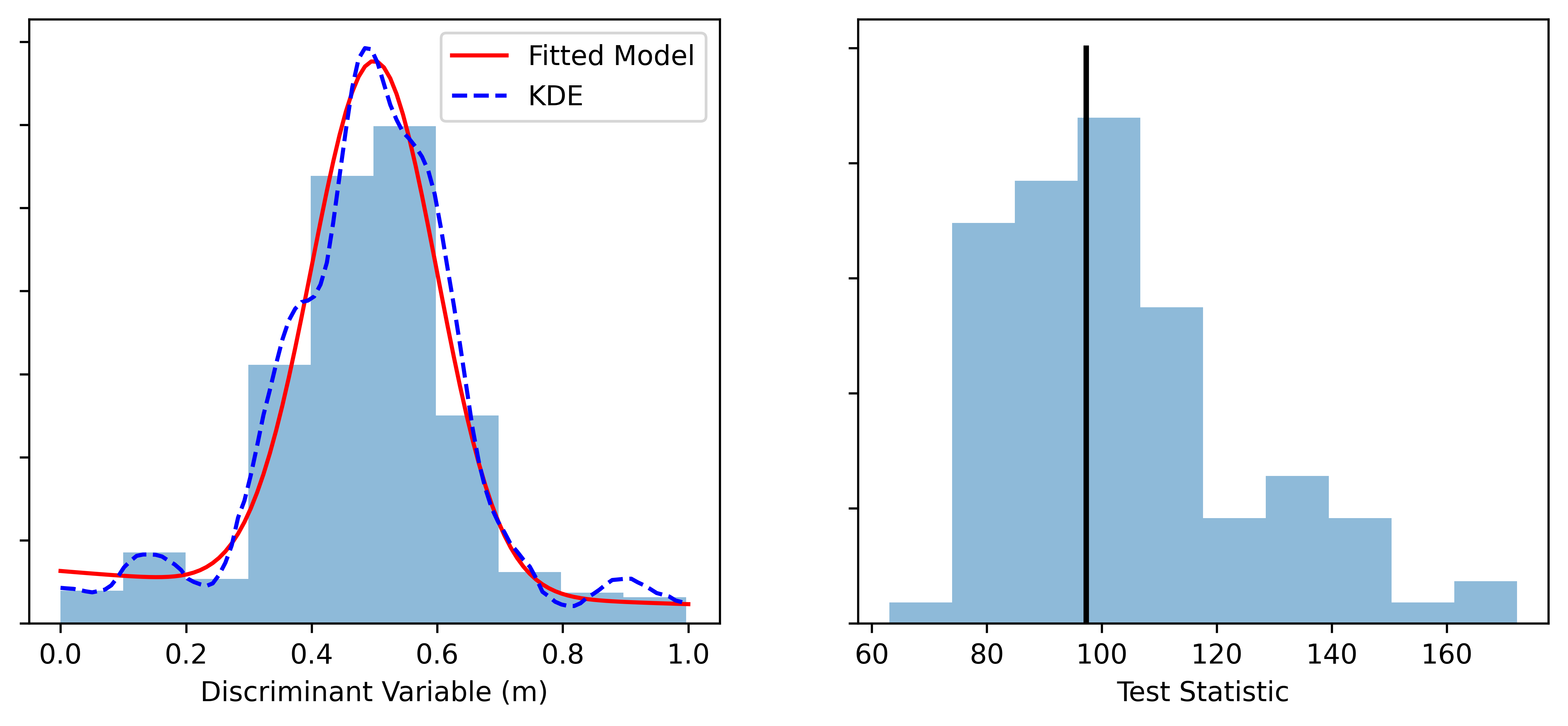}
\caption{Aspects of a goodness-of-fit
test. The left panel shows the two fits (the fitted model being
assessed and a kernel density estimate), at the slice centered on $t=0.4125$. The right panel shows the
distribution of the test statistic under the null hypothesis,
simulated using a semiparametric bootstrap procedure. The
vertical line is the position of the observed test statistic.}
\label{fig::gofmethod}
\end{figure}

Using this as a starting point, a goodness-of-fit test
can be simply constructed by choosing an evenly-spaced
grid of values $t_1, t_2,\dots, t_K$. (Here, $K=20$ was used.) For each $k$, observations were weighted using a Gaussian
kernel which measures similarity to $t_k$ in the $T$ direction.
Models are fit to the weighted sample via two approaches.
First, the maximum likelihood fit to determine the best-fitting
mixture of the $g$'s, and, second, a kernel density estimate.
The left panel of Figure \ref{fig::gofmethod} shows these
two fits for one such slice.

The test statistic used in making this comparison is the
L2 distance between the two fits, summed across all $K$
slices. The distribution of this test statistic
under the null hypothesis that the chosen model is
correct is approximated using a semiparametric bootstrap
procedure in which values of $T$ are first sampled with
replacement from the observed data. For each sampled
value, the hypothesized model is used to simulate a
value for $M$. A sample of size equal to the original
sample is generated in this way, and the entire fitting
procedure is repeated for this sample, with the calculated
test statistic retained. The right panel of Figure
\ref{fig::gofmethod} shows the distribution of the
simulated test statistics compared with the observed
value. Clearly, in this case there is not evidence to
conclude that the model is not a good fit.

\subsection{Asymptotic Distribution and Confidence Band}

Likewise, a simple procedure exists for constructing
a confidence band on the resulting fit. For
simplicity, assume $s=b=1$, but the result can be
extended without much difficulty.
Let
$f(t) = z h_1(t)$ and
consider the 
weighted kernel estimate 
$$
\hat f(t) = \sum_i w_1(M_i) K_\nu(T_i-t)
$$
with bandwidth $\nu$.
Define $\hat z = \int \hat f(t) dt$
and 
$\hat h_1(t) = \hat f(t) / \hat{z}.$
By standard calculations,
$\E[\hat h_1(t)] = h_1(t) + O(\nu^2)$,
$\mathrm{Var}[\hat h_1(t)] = O(1/(n\nu))$
and if $\nu\to 0$, $n\nu\to\infty$, $\nu = o(1/n^{1/5})$,
then
$$
\sqrt{nh} (\hat h_1(t)-h_1(t))\rightsquigarrow N(0,\sigma^2(t)).
$$
A confidence band can be obtained by the bootstrap
\citep{chernozhukov2014} as follows.
Draw $B$ bootstrap replications yielding
estimates $\hat h_{1j}^*$ for $j=1,\ldots, B$.
Let $c_\alpha$ be the $1-\alpha$ quantile of
$\sup_t |\hat h_{1j}^*(t) - \hat h_{1}(t)|$.
Then
$\hat h_1(t) \pm c_\alpha$ gives an asymptotic
$1-\alpha$ confidence band.

\subsection{The Herd: Identifiability of COWs}

In this section,
we investigate the identifiability
of the COWs model.
We again begin with the sPlot case where $s=b=1$.
From \cite{allman2009, hall2003}
we know that this model is not identified.
There are infinitely many choices of
$z, g_1,g_2,h_1,h_2$ that give
the same joint density $p(m,t)$.
However, in the sPlot framework we also
assume that $g_1$ and $g_2$ are known
(or can be separately estimated).
In this case, the model is identified
and we can consistently estimate
$z, h_1, h_2$.
As mentioned above,
if there were three or more variables ---
rather than just $M$ and $T$ ---
then the model is fully identified.
In Section \ref{section::extensions},
we discuss an extension of sPlot
which does not require $g_1$ and $g_2$ to be known or
estimated from external data.

Now consider the COWs model
\begin{equation}\label{eq::again}
\p(m,t) = \sum_{j=1}^s z_{j}g_{j}(m) h_{j}(m) + 
\sum_{j=s+1}^{s+b} z_{j}g_{j}(m) h_{j}(m),
\end{equation}
and the most favorable situation
where the background $\back(m)$ and
signal $\sig(m)$  for $M$ are known.
The COWs model
splits the background $\back$ and signal $\sig$
into sums of functions:
\begin{equation}\label{eq::restrict}
\sig(m) \propto \sum_{j=1}^s z_{j}\,g_{j}(m),\ \ \ 
\back(m) \propto \sum_{j=s+1}^{b+s} z_{j}\, g_{j}(m),
\end{equation}
where $\int \sig(m) dm = \int \back(m) dm = 1$.
This has the effect of
relaxing the conditional independence assumption.
But there are infinitely many ways to
split the functions.
Is this choice identified?

In fact, there are two different types of non-identifiability that can occur
and one is harmless while the other is problematic.
It is easy to see using the results of Theorem 4.1 of \cite{hall2003}
that we can rewrite this as a different decomposition
$$
\p(m,t) = \sum_{j=1}^s \tilde z_j \tilde g_{j}(m)\tilde h_{j}(t) + 
\sum_{j=s+1}^{s+b} \tilde z_j \tilde g_{j}(m)\tilde h_{j}(t).
$$
However, 
the marginal signal density
is preserved:
$$
\sig(t)= \frac{\sum_{j=1}^s  z_j h_{j}(t)}{\sum_{j=1}^s z_j}=
\frac{\sum_{j=1}^s \tilde z_j \tilde h_{j}(t)}{\sum_{j=1}^s \tilde z_j}
$$
and so our conclusions about
the signal in $T$ would be unchanged.
Given $\p(m,t)$, $\back(m)$, $\sig(m)$, $s$ and $b$,
let
${\cal C}(\p,\sig,\back,s,b)$
denote all models of the form
(\ref{eq::again}).
We call 
${\cal C}(\p,\sig,\back,s,b)$ the herd of COWs models.
The key identifiability question is the following:
does every model in 
${\cal C}(\p,\sig,\back,s,b)$
have the same signal density
$\sig(t)$?
Currently this question is open.

\subsection{COWs without Weights via Least Squares}

In this section, we 
show that the COWs model can also be estimated
using least squares if we have access to an estimate of the joint distribution of $M$ and $T$.
Suppose, first, that we bin the data
and represent $p(m,t)$
as a matrix {\bf P}
of dimension $n_m \times n_t$.
The functions
${\bf g}_1,{\bf g}_2\ldots, {\bf g}_{s+b}$ are now vectors of length $n_m$ and likewise the ${\bf h}_j$
are vectors of length $n_t$.
Hence,
$$
{\bf P} = \sum_{j=1}^{s+b} z_j {\bf g}_j {\bf h}_j^T.
$$
Let ${\bf G}$ 
be the $n_m\times N$ matrix whose
$j$-th column is ${\bf g}_j$.
Let ${\bf H}$ be the $n_t\times N$ matrix whose
$j$-th column is $z_j {\bf h}_j$
so that we have absorbed the $z_j$ in ${\bf H}$.
Thus we can write
$$
{\bf P} = {\bf G} {\bf H}^T.
$$
Let $\widehat {\bf P}$ be the histogram estimate of ${\bf P}$.
The least squares estimate of ${\bf H}$ minimizes
$$
||\widehat {\bf P} - {\bf G} {\bf H}^T ||^2.
$$
The solution is
$$
\widehat {\bf H}^T = ({\bf G}^T {\bf G})^{-1} {\bf G}^T \widehat {\bf P}.
$$
This corresponds exactly to the COWs solution using the weight matrix
$$
{\bf W} = {\bf G} ({\bf G}^T {\bf G})^{-1}.
$$
Define ${\bf e} = (1,\ldots, 1,0,\ldots, 0)^T \in \mathbb{R}^N$,
where $e_j = I(j\leq s)$.
The estimate of the signal vector $\sig$ is then
$$
\widehat \sig = \widehat {\bf H} {\bf e} = \widehat {\bf P}^{T} {\bf G} ({\bf G}^T {\bf G})^{-1}{\bf e}.
$$

There is an alternative approach that avoids binning.
Let
$\widehat p(m,t)$ be an estimate of the density $\p(m,t)$,
such as a kernel density estimate.
For convenience, define
$r_j(t) = z_j h_j(t)$.
Fix a value $t$.
We then have
$$
\hat p(M_i,t) \approx 
r_1(t) g_1(M_i) + \ldots + r_N(t) g_N(M_i)
$$
which is simply a linear regression equation.
We choose
$\widehat {\bf r}(t) = (\widehat r_1(t),\ldots, \widehat r_N(t))^T$ to minimize
$\| {\bf Y}(t) - {\bf G} \widehat {\bf r}(t) \|^2$,
where ${\bf Y}(t) = (\widehat p(M_1,t),\ldots, \widehat p (M_n,t))^T$ and
${\bf G}$ is the $n\times N$ matrix with $(i,j)$ entry
$g_j(M_i)$.
The solution is
$$
\widehat {\bf r}(t) = ({\bf G}^T {\bf G})^{-1} {\bf G}^T {\bf Y}(t).
$$
Then
$$
\widehat \sig(t) \propto \sum_{j=1}^s \hat r_j(t).
$$

\section{Mixtures and Copulas}\label{section::copula}

The original sPlot model
is a mixture model which
assumes that $M$ and $T$ are independent
within the signal and background.
That is, it assumes that $M$ and $T$ are independent
given $S$, where
$S\in \{1,2\}$ denotes background or signal.
The COWs model
relaxes this conditional independence assumption.
In this section,
we describe a different way
to relax the assumption of conditional independence
using a mixture of copulas.
Recall that a copula $C$ is a distribution on
$[0,1]^2$ with uniform marginals.
We let $c$ denote the density of the copula.
By Sklar's theorem \citep{sklar1959}, any joint density
$f(x,y)$ 
for two random variables $X$ and $Y$ can be written as
$$
f(x,y) = f_X(x)f_Y(y) c(F_X(x),F_Y(y))
$$
for some copula density $c$,
where
$F_X$ and $F_Y$ are the cdf's of $X$ and $Y$ with densities
$f_X$ and $f_Y$.

We propose to model the joint density in $M$ and $T$ as the following mixture of copulas:
\begin{equation}\label{eq::mixcop}
\p(m,t) = z g_1(m)h_1(t)c_1(G_1(m),H_1(t);\theta_1) + (1-z) g_2(m)h_2(t)c_2(G_2(m),H_2(t);\theta_2).
\end{equation}
This is a semiparametric model with parameters
$z, \theta_1, \theta_2, h_1(\cdot),\: \mbox{and}\:h_2(\cdot).$
We continue to assume that $g_1$ and $g_2$
are known or can be estimated
from auxiliary data
or from the marginal of $M$, otherwise the model is not identified.
To be concrete, we will use the Gaussian copula, but
similar ideas can be applied to other choices.
A distribution of a vector ${\bf X}$ 
is a Gaussian copula if
$f({\bf X}) \sim N(0,{\bf R})$
where
$f({\bf x}) = (f_1(x_1),\ldots, f_d(x_d))$
is a vector of monotonic transformations
and ${\bf R}$ is a correlation matrix.

Estimating the parameters of a single Gaussian copula is straightforward.
The function $f_j$ can be estimated by
$$
\hat f_j(x_j) = \Phi^{-1}(\hat F_j(x_j)),
$$
where $\Phi$ is the CDF of a standard normal and
$$
\hat F_j(x_j) = \frac{\sum_i I(X_{ij} \leq x_j)}{n+1}.
$$
Then we estimate ${\bf R}$ by computing the sample
correlation matrix $\widehat {\bf R}$ of the transformed data
$\hat f(X_1),\ldots, \hat f(X_n)$.
Theoretical properties of such estimators
are described in 
\cite{segers2014}, \cite{hoff2014}, and \cite{chen2006}.

\begin{figure}
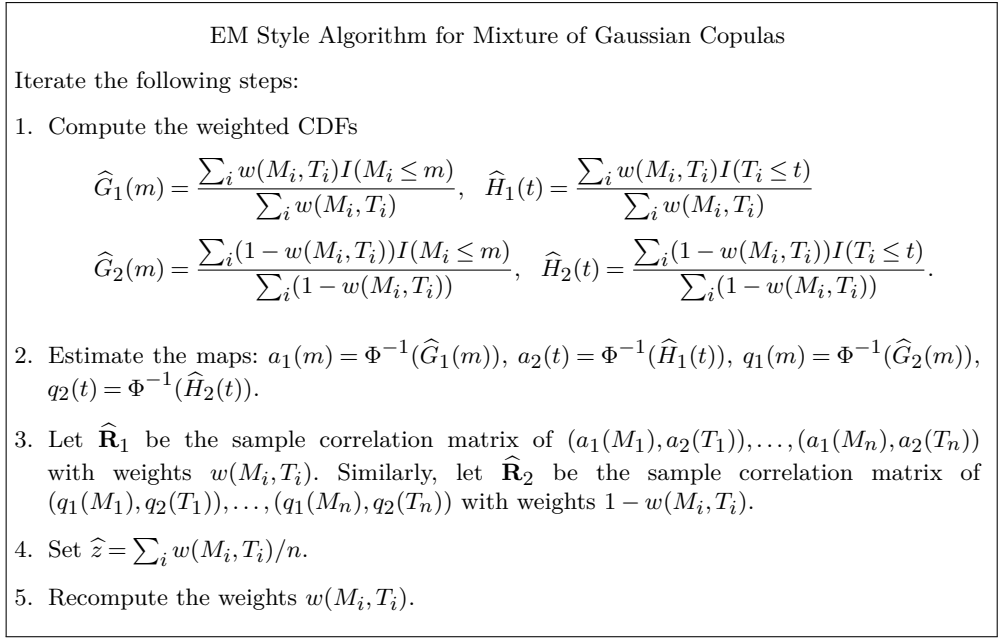

\fbox{\parbox{0.9\textwidth}{
\begin{center}
EM Style Algorithm for Mixture of Gaussian Copulas
\end{center}

Iterate the following steps:

\begin{enumerate}
\item Compute the weighted CDFs
\begin{align*}
\hat G_1(m) &= \frac{\sum_i w(M_i,T_i) I(M_i \leq m)}{\sum_i w(M_i,T_i)},\ \ 
\hat H_1(t) = \frac{\sum_i w(M_i,T_i) I(T_i \leq t)}{\sum_i w(M_i,T_i)}\\
\hat G_2(m) &= \frac{\sum_i (1-w(M_i,T_i)) I(M_i \leq m)}{\sum_i (1-w(M_i,T_i))},\ \ 
\hat H_2(t) = \frac{\sum_i (1-w(M_i,T_i)) I(T_i \leq t)}{\sum_i (1-w(M_i,T_i))}.
\end{align*}

\vspace{.1in}
\item Estimate the maps:
$a_1(m) = \Phi^{-1}(\hat G_1(m))$,
$a_2(t) = \Phi^{-1}(\hat H_1(t))$,
$q_1(m) = \Phi^{-1}(\hat G_2(m))$,
$q_2(t) = \Phi^{-1}(\hat H_2(t))$.

\vspace{.1in}
\item Let $\widehat {\bf R}_1$ be the sample correlation matrix
of $(a_1(M_1),a_2(T_1)),\ldots, (a_1(M_n),a_2(T_n))$
with weights $w(M_i,T_i)$.
Similarly,
let $\widehat {\bf R}_2$ be the sample correlation matrix
of $(q_1(M_1),q_2(T_1)),\ldots, (q_1(M_n),q_2(T_n))$
with weights $1-w(M_i,T_i)$.

\vspace{.1in}
\item Set
$\hat z = \sum_i w(M_i,T_i)/n$.

\vspace{.1in}
\item Recompute the weights $w(M_i,T_i)$.
\end{enumerate}
}
}

\caption{\em Estimation algorithm for mixture of copulas.}
\label{fig::copula}
\end{figure}

This mixture of copulas model can be estimated using an
EM approach, as in 
\cite{levine2011} and \cite{chauveau2015}.
Our algorithm is simplified by our use
of a Gaussian copula.
For the first copula, there is
a transformation
$a=(a_1,a_2)$ such that
$(a_1(M),a_2(T))\sim N(0,{\bf R}_1)$.
For the second copula, there is
transformation $q=(q_1,q_2)$ such that
$(q_1(M),q_2(T))\sim N(0,{\bf R}_2)$.
Let
$$
w(M,T) = P(S=1|M=m,T=t) =
\frac{z g_1(m) h_1(t) c(G_1(m),H_1(t);{\bf R}_1)}{p(m,t)}
$$
which gives the weights to be used in the algorithm.
The algorithm is given in Figure \ref{fig::copula}.


\begin{figure}
\begin{center}
\begin{tabular}{cc}
\includegraphics[scale=.35]{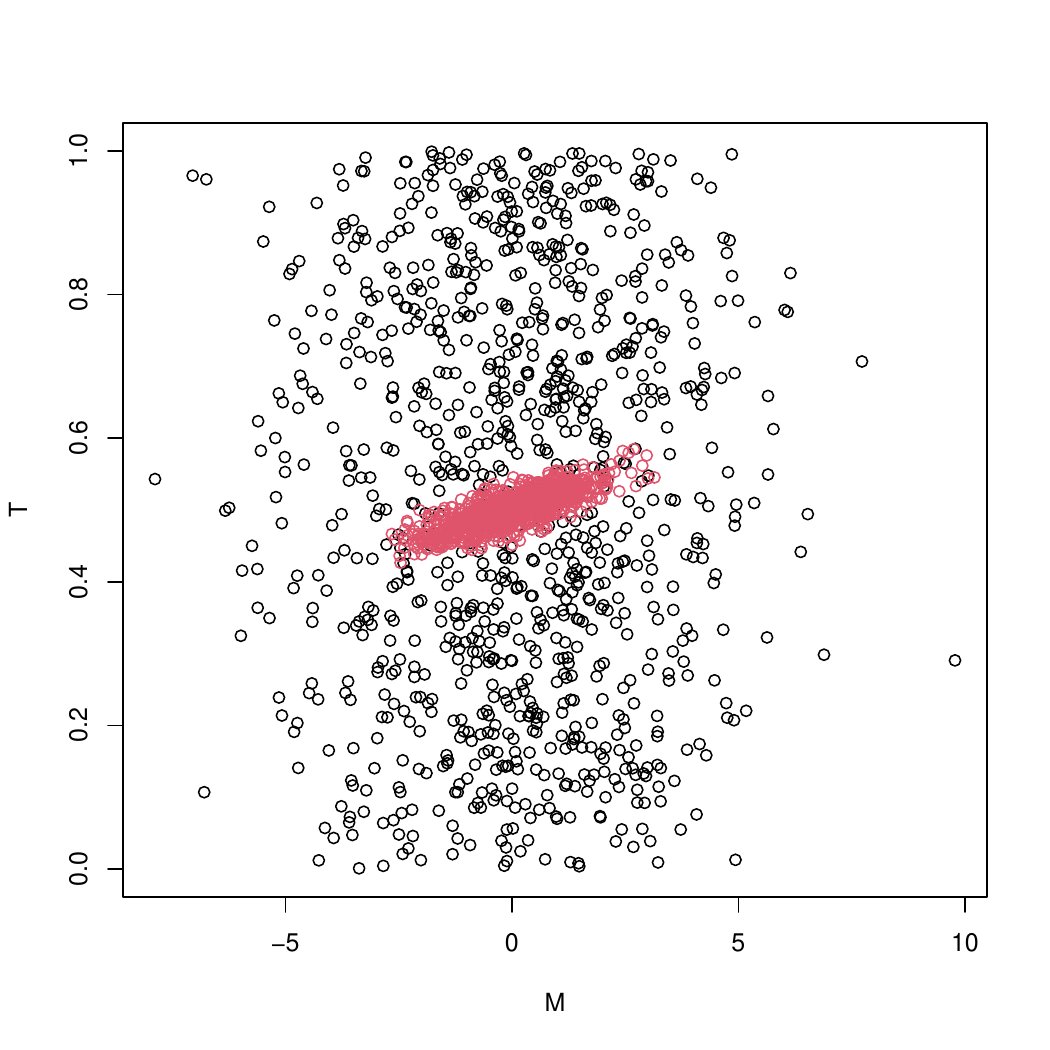} &
\includegraphics[scale=.35]{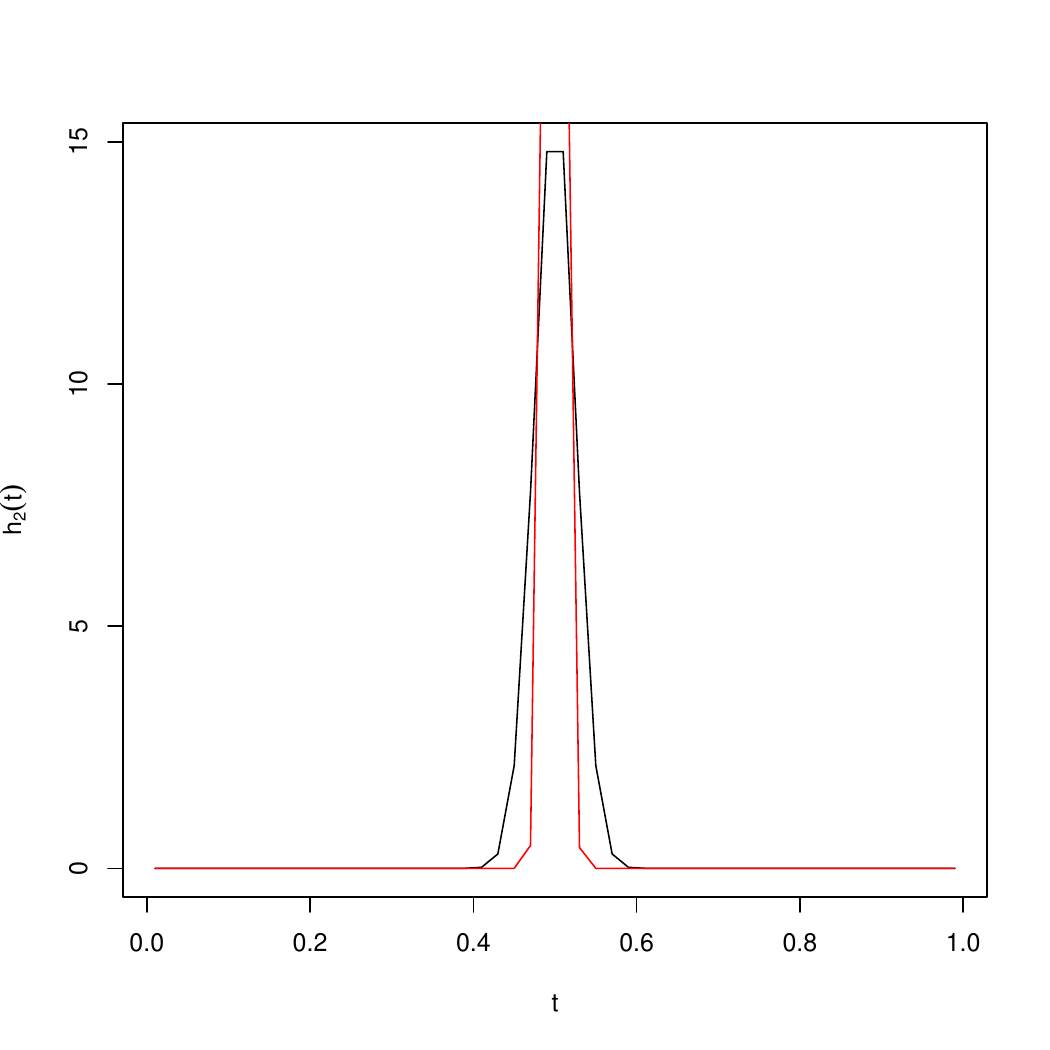} 
\end{tabular}
\end{center}
\caption{Left: data. Right: estimate (black) and true density $h_1(t)$
for the copula example.}
\label{fig::copuladat}
\end{figure}

This algorithm gives
$\hat z$, $\widehat {\bf R}_1$ and $\widehat {\bf R}_2$
and estimated maps
$a$ and $q$.
Since $a_2(T)\sim N(0,1)$,
the signal density is
$h_1(t) = \phi(a_2(t))\ |a_2'(t)|$,
where $\phi$ is the density of a standard normal.
Hence, we take
$\hat h_1(t) = \phi(\tilde a_2(t))\ |\tilde a_2'(t)|$,
where $\tilde a_2$ is a smoothed version of $a_2$.
Alternatively, we can use a kernel estimator
for $T_1,\ldots, T_n$
with weights from the final step of the EM algorithm.
Figure \ref{fig::copuladat}
shows some data from the model
from a mixture of Gaussian copulas.
The right plot shows the true density $h_1(t)$ in red
and the estimated density.

\subsection{Relating COWs and Copulas}
The COWs model is, in fact, a mixture of
copulas.
To see this,
write the marginals for the background and signal as
\begin{eqnarray*}
\sig(m) & = & \sum_{j=1}^s z_j g_j(m)\bigg/\sum_{j=1}^s z_j \\
\sig(t) & = & \sum_{j=1}^s z_j h_j(m)\bigg/\sum_{j=1}^s z_j \\
\back(m) &=& \sum_{j=s+1}^{b+s} z_j g_j(m)\bigg/\sum_{j=s+1}^{b+s} z_j \\
\back(t) &=& \sum_{j=s+1}^{b+s} z_j h_j(m)\bigg/\sum_{j=s+1}^{b+s} z_j
\end{eqnarray*}
Denote the corresponding CDFs by
$\sigCDF(m)$, $\sigCDF(t)$, $\backCDF(m)$ and $\backCDF(t)$.
Let $\lambda = \sum_{j=1}^s z_j$.
The COWs model can then be written as
$$
p(m,t) =  
\lambda \sig(m) \sig(t) 
c_{1}(\sigCDF(m),\sigCDF(t)) +
(1-\lambda) \back(m) \back(t) 
c_{2}(\backCDF(m),\backCDF(t))
$$
where
$$
c_{1}(u,v)=
\frac{\sum_{j=1}^s z_j g_j(\sigCDF_M^{-1}(u)) h_j(\sigCDF_T^{-1}(v))}
{\Biggl(\sum_{j=1}^s z_j g_j(\sigCDF_M^{-1}(u)) \Biggr) \Biggl(\sum_{j=1}^s z_jh_j(\sigCDF_T^{-1}(v))\Biggr)}
$$
and
$$
c_{2}(u,v)=
\frac{\sum_{j=s+1}^{b+s} z_j g_j(\backCDF_{M}^{-1}(u)) h_j(\backCDF_{T}^{-1}(v))}
{\Biggl(\sum_{j=s+1}^{b+s} z_j g_j(\backCDF_{M}^{-1}(u)) \Biggr) \Biggl(\sum_{j=s+1}^{b+s} z_j h_j(\backCDF_{T}^{-1}(v))\Biggr)}.
$$
Different choices of $g_j$
give different copulas.
This seems to be a rather complex copula.
It would be interesting to understand
how the COWs copula relates to more traditional copulas. A special case where this connection is clear is sPlot with $s=b=1$ in which case the COWs copula reverts to the independence copula, i.e., $c_1(u,v) = c_2(u,v) = 1$.

The COWs model 
was not intended to be a copula.
Nonetheless, the model is interesting
because it is easy to fit compared to
most mixtures of copulas.
Instead of thinking of the $g_j$'s as basis functions for $M$,
we can instead treat the $g_j$'s
as a tool for
defining a copula.
Doing so will allow us to directly construct
flexible copulas that relax the conditional independence
assumption.
And the resulting mixture can be fit by least squares
which is in sharp contrast
to the existing mixtures of copulas.

{\bf Remark.}
We have assumed that the copulas $c_1$ and $c_2$
are parametric.
If we made $c_1$ and $c_2$
completely nonparametric copulas,
the model may no longer be identified.
But, it might be possible to use
nonparametric copulas if we added shape constraints.

\section{Some Other Extensions of sPlot and COWs}
\label{section::extensions}

We have seen that mixtures of copulas
are one example of a useful extension of sPlot.
In this section, we 
consider other extensions.

\subsection{Leveraging sPlot For General Conditional Independence Mixtures}
\label{section::leverage}

The general mixture of conditional independence distributions model is
\citep{hall2003, allman2009, bonhomme2016}
$$
p(x) =p(x_1,\ldots, x_d)= \sum_{j=1}^k \pi_j \prod_{r=1}^d f_{r}^{(j)}(x_r)
$$
for a random vector
${\bf X} = (X_1,\ldots, X_d)$.
This is a mixture model
in which the coordinates of ${\bf X}$
are conditionally independent within each
component of the mixture.
As mentioned earlier,
it has been shown that this model is
fully nonparametrically identified
when $d\geq 3$, which we shall assume throughout this section.
There are two classes of algorithms for estimation in this model.
The first are EM-type methods
\citep{levine2011};
the second are matrix and tensor optimization methods 
\citep{bonhomme2016}.
Here we show how sPlot 
can be used to develop a new, simple algorithm.
The idea is as follows: treat one dimension as fixed and known,
find the sPlot weights, estimate the densities of the other dimensions,
now cycle through all the dimensions.

First, recall that given $k$ densities
$f^{(1)},\ldots, f^{(k)}$, it is possible to find $k$ weight functions
$w^{(1)},\ldots,w^{(k)}$ such that
$\int f^{(j)} w^{(\ell)} = 1$ if $j=\ell$ and 0 otherwise.
Specifically, define
the $k\times k$ matrix ${\bf F}$ with
$F_{ij} = \int f^{(i)}(x)f^{(j)}(x) dx$
and let ${\bf A} = {\bf F}^{-1}$.
Then we take
$$
w^{(j)} = A_{j1}f^{(1)} + \cdots + A_{jk}f^{(k)}.
$$
Define an operator $\Gamma$
that takes in $f^{(1)},\ldots,f^{(k)}$
and outputs the weight functions
$$\Gamma(f^{(1)},\ldots, f^{(k)}) = \{w^{(1)},\ldots,w^{(k)}\}.$$
Now iterate the following steps over $r=1,2,\ldots, d$:

\begin{enumerate}
\item Treat
$\hat f_r^{(1)},\ldots,\hat f_r^{(k)}$ as known
and let
$(w_r^{(1)},\ldots,w_r^{(k)}) = \Gamma(\hat f_r^{(1)},\ldots,\hat f_r^{(k)})$
be the weight functions.

\item For the coordinates $s\neq r$ and $j=1,\ldots, k$,
let
$$
\hat f_s^{(j)}(x_s) =
\frac{\sum_i w_{r}^{(j)}(X_{si})K_h(x_s - X_{si})}
{\sum_i w_{r}^{(j)}(X_{si})}
$$
where
$X_{si}$ denotes the $i^{\rm th}$ observation of $X_s$.

\item
Let
$$
\hat \pi_j = \int \cdots \int w_1^{(j)}(x_1) \cdots w_d^{(j)}(x_d) \,dx_1 \cdots dx_d.
$$
\end{enumerate}

This is as simple as the EM algorithm
but is far simpler than the other algorithms.
It is solving a least squares problem
while EM solves the MLE.
That suggests it might be more robust to outliers.

\subsection{Localized Signals}

Consider again the sPlot model but drop
the independence assumption.
Thus
$$
p(m,t) = z g_1(m)h_1(t|m) + (1-z) g_2(m)h_2(t|m) =
z b(m,t) + (1-z) s(m,t),
$$
where
$g_1,g_2,z$ are taken as known.
In some cases, it is assumed that the signal is localized.
Suppose then that
the signal for $M$ is localized to an interval
$[c_1,c_2]$.
For $m\notin [c_1,c_2]$, there is no signal
so we can estimate the joint background
$\back(m,t)$. Hence, we can also estimate
$\back(t|m)$ for $m\notin [c_1,c_2]$.
We do not have access to
$\back(t|m)$ in the signal region $c_1 < m < c_2$.
Instead, we shall interpolate $\back(t|m)$ using optimal transport.
Let us briefly recall some detail about optimal transport.
(See \cite{villani2021} for more details.)

The optimal transport map from $P$ to $Q$ is the map $T$ that minimizes
$\int ||T(x)-x||^2 dP(x)$ subject to the condition that
$X\sim P$ implies that $T(X)\sim Q$.
If $X$ is one dimensional, $P$ has a density,
and $F$ and $G$ are the cdf's of $P$ and $Q$, then
$T(x) = G^{-1}(F(x))$.
The minimum value of
$\int ||T(x)-x||^2 dP(x)$ 
is the squared Wasserstein distance denote by
$W^2(P,Q)$.
There is a shortest path of distributions
$(P_t:\ 0 \leq t \leq 1)$
connecting $P$ and $Q$ called the
Wasserstein geodesic
where $P_0=P$ and $P_1 = Q$.
The distribution $P_t$ has cdf
$F_t$ given by
$$
F_t^{-1}(u) = (1-t)F^{-1}(u) + t G^{-1}(u).
$$

In our setting,
we let $\back(t|m)$ be the Wasserstein geodesic
connecting $\back(t|c_1)$ to $\back(t|c_2)$.
This $\back(t|m)$ has a cdf $\backCDF(t|m)$ whose inverse is given by
$$
\backCDF^{-1}(u|m) = \frac{m-c_1}{c_2-c_1} \backCDF^{-1}(u|c_2) +
\frac{c_2 - m}{c_2-c_1} \backCDF^{-1}(u|c_1).
$$
This completely determines $\backCDF(t|m)$ and hence $\back(t|m)$.
Now we have $\back(t|m)$ for all $m$ and $t$.
Hence
$\back(m,t) = g_1(m)\back(t|m)$
and
the joint signal density is
$$
\sig(m,t) = \frac{p(m,t)-z \back(m,t)}{1-z}.
$$
Everything on the right-hand side can be estimated, so
$$
\hat \sig(m,t) = \frac{\hat p(m,t)-\hat z \, \hat \back(m,t)}{1-\hat z}
$$
and
$$
\widehat \sig(t) = \int \widehat \sig(m,t) dm.
$$
It is possible to derive the limiting distribution 
of $\widehat \sig(t)$ from which we can get confidence bands
but we do not pursue that here due to space
limitations.

\smallskip

\begin{figure}
\begin{center}
\includegraphics[scale=.5]{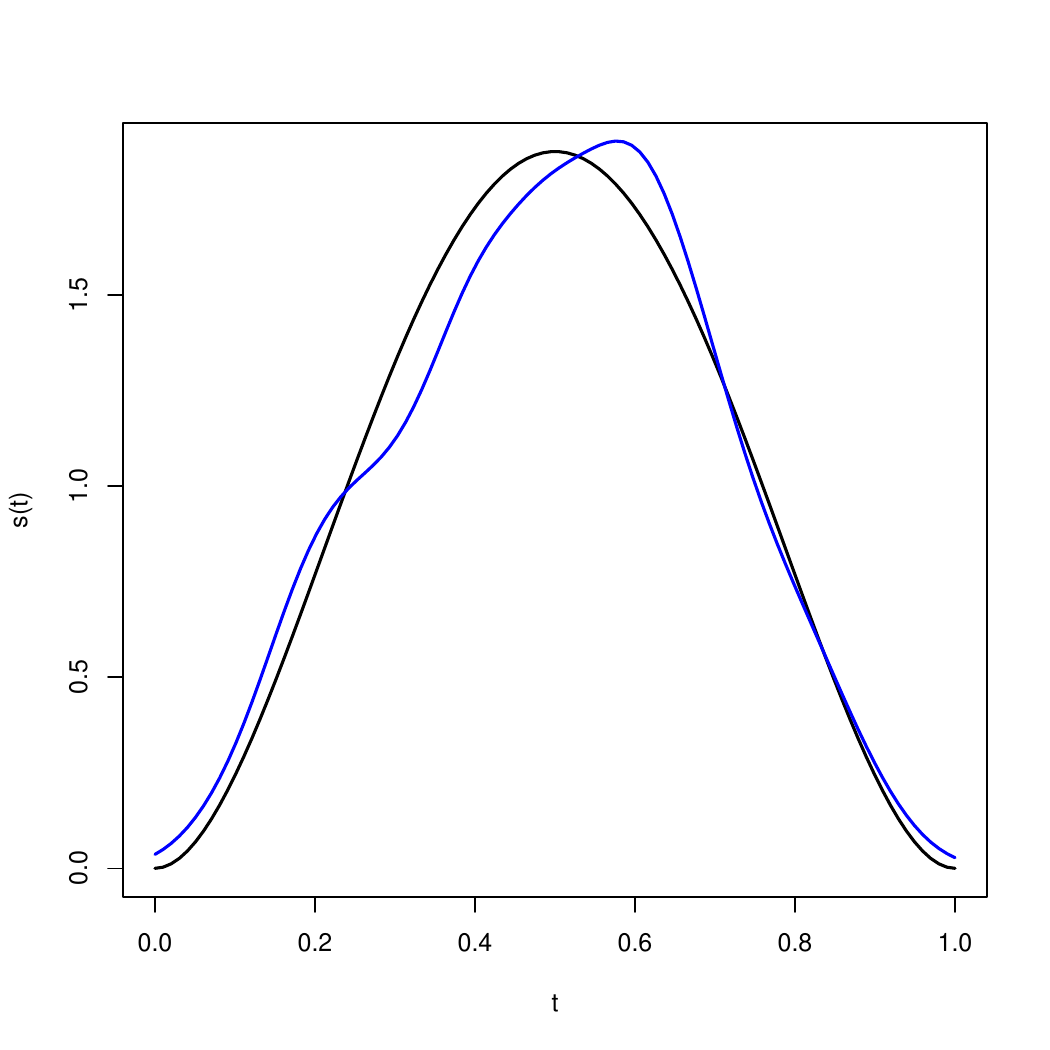}
\end{center}
\caption{\em True signal density $\sig(t)$ (black curve) and estimated density
$\widehat \sig(t)$ (blue curve) using the optimal transport interpolation method.}
\label{fig::OT}
\end{figure}

{\bf Example.}
As a proof of concept,
we generated 1000 observations
from
$\p(m,t) = (1/2) \back(m,t) + (1/2) \sig(m,t)$,
where 
$\back(m)$ is Unif(0,1),
$\back(t|m) = \mathrm{Beta}(1+4m,5-4m)$, 
$\sig(m,t)=\sig(m)\sig(t)$, where
$\sig(m)$ is a Beta(3,3) rescaled to $[.4,.6]$ and
$\sig(t)$ is Beta(3,3).
The true and estimated densities are shown in Figure
\ref{fig::OT}.
The estimate of $\back(t|m)$ was obtained by performing
local linear regression of
$K_h(T_i-t)$ on $M_i$,
where $K$ is a normal kernel and the bandwidth was
$h=0.05$.

\smallskip

We used optimal transport to interpolate
$\back(t|m)$ but other methods
can also be used.
For example,
\cite{sengupta2024} and \cite{raine2023}
use neural networks and normalizing flows.
These seem to work well but they are
very black box in nature and it is not easy to
derive their statistical properties.

\subsection{sPlot With No Assumptions}

Although $h_1(t)$ is not identified in the sPlot model,
we now show it is possible to put bounds on it.
\cite{hall2003} showed that
the set ${\cal A}$ of all 
$(\tilde z, \tilde g_1(m), \tilde h_1(t), \tilde g_2(m), \tilde h_2(t))$
that give valid decompositions
$p(m,t) = \tilde z \, \tilde g_1(m) \tilde h_1(t) + (1-\tilde z)\tilde g_2(m) \tilde h_2(t)$
is a two-parameter family 
that we describe below.
However,
the labels 1 and 2 are not identified.
We can always permute 1 and 2 and get another solution.
To resolve this indeterminancy, we make one mild assumption:
we assume that the signal $g_1(m)$ is more concentrated
than the background $g_2(m)$.
Thus we assume that
$\psi(g_1) \leq \psi(g_2)$
where $\psi$ is some functional that
measures how concentrated the density is.
We use the entropy
$\psi(g) = -\int g(m)\log g(m) dm$,
although other measures can be used.
We define
$$
{\cal A}^\dagger =
\Bigl\{
(\tilde z, \tilde g_1(m), \tilde h_1(t), \tilde g_2(m), \tilde h_2(t))\in {\cal A}:\
\psi(\tilde g_1) \leq \psi(\tilde g_2)
\Bigr\}.
$$
Let
${\cal H}_1$ be the set of all $\tilde h_1$ in ${\cal A}^\dagger$.
Then we define
$$
\ell(t) = \inf_{h_1\in {\cal H}_1}h_1(t),\ \ \
u(t) = \sup_{h_1\in {\cal H}_1}h_1(t).
$$

Now we describe the
two parameter family of decompositions.
Suppose we have an initial decomposition
$p(m,t) = z g_1(m)h_1(t) + (1-z)g_2(m) h_2(t)$.
Let
\begin{align*}
w_1(m) &= \sqrt{z (1-z)} (g_1(m)-g_2(m))\\
w_2(t) &= \sqrt{z (1-z)} (h_1(t)-h_2(t)).
\end{align*}
For each pair of real numbers
$(\alpha_1,\beta_1)$ such that
$\mathrm{sign}(\alpha_1) = - \mathrm{sign}(\beta_1)$, let
$\alpha_2 = -1/\beta_1$, $\beta_2 = -1/\alpha_1$ and let
$\tilde z = |\beta_1|/(|\beta_1| + |\alpha_1|)$ and
\begin{align*}
\tilde g_1(m) &= p(m) + \alpha_1 w_1(m),\ \ \ \tilde g_2(m) = p(m) + \beta_1 w_1(m)\\
\tilde h_1(t) &= p(t) + \alpha_2 w_2(t),\ \ \ \tilde h_2(t) = p(t) + \beta_2 w_2(t).
\end{align*}
Let
$$
\Theta = \Bigl\{(\alpha_1,\beta_1):
\inf \tilde g_1(m) \geq 0,
\inf \tilde g_2(m) \geq 0,
\inf \tilde h_1(t) \geq 0,
\inf \tilde h_2(t) \geq 0\Bigr\}.
$$
Let
${\cal A}$
be the set of all
$(\tilde z, \tilde g_1, \tilde g_2, \tilde h_1, \tilde h_2)$
of this form,
where $(\alpha_1,\beta_1) \in\Theta$.
This defines the valid decompositions of
$p(m,t)$.

To get an initial solution,
we compute $p(m,t)$ on a grid and apply
nonnegative matrix factorization (see Section~\ref{sec:NMF}).
This will give
$e_1,e_2,f_1,f_2$ such that
$$
p(m,t) = e_1(m)f_1(t) + e_2(m)f_2(t) .
$$
We can write this as
$$
p(m,t) = z g_1(m)h_1(t) + (1-z)g_2(m) h_2(t)
$$
where
$$
g_1(m) = \frac{e_1(m)}{\int e_1(m) dm},\ 
g_2(m) = \frac{e_2(m)}{\int e_2(m) dm},\ 
h_1(t) = \frac{f_1(t)}{\int f_1(t) dt},\ 
h_2(t) = \frac{f_2(t)}{\int f_2(t) dt}
$$
and
$z = \int e_1(m)dm \int f_1(t)dt$.
This defines the initial solution.
In practice,
we first get an estimate
$\hat p(m,t)$ of the joint density $p(m,t)$,
then we apply the above method.

\begin{figure}
\begin{center}
\begin{tabular}{cc}
\includegraphics[scale=.35]{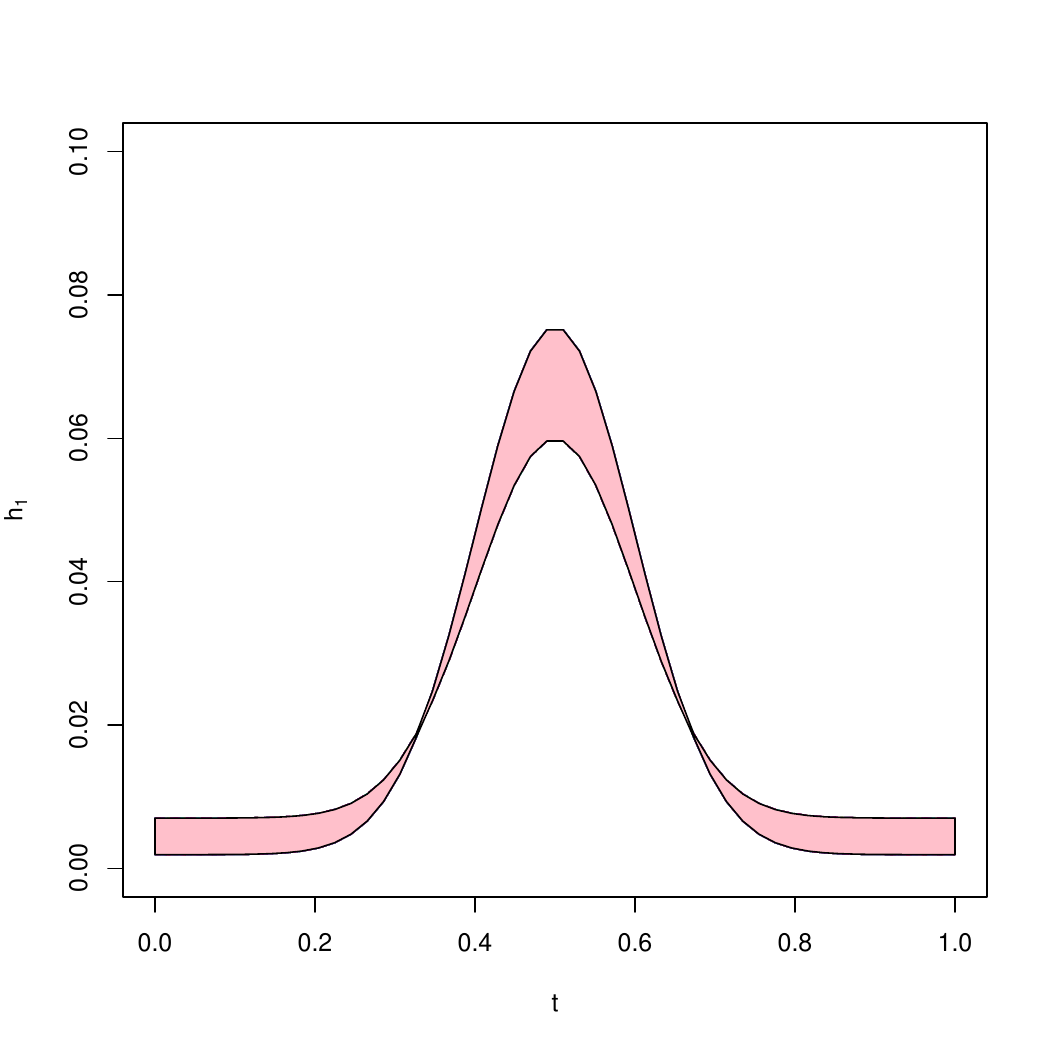} &   
\includegraphics[scale=.35]{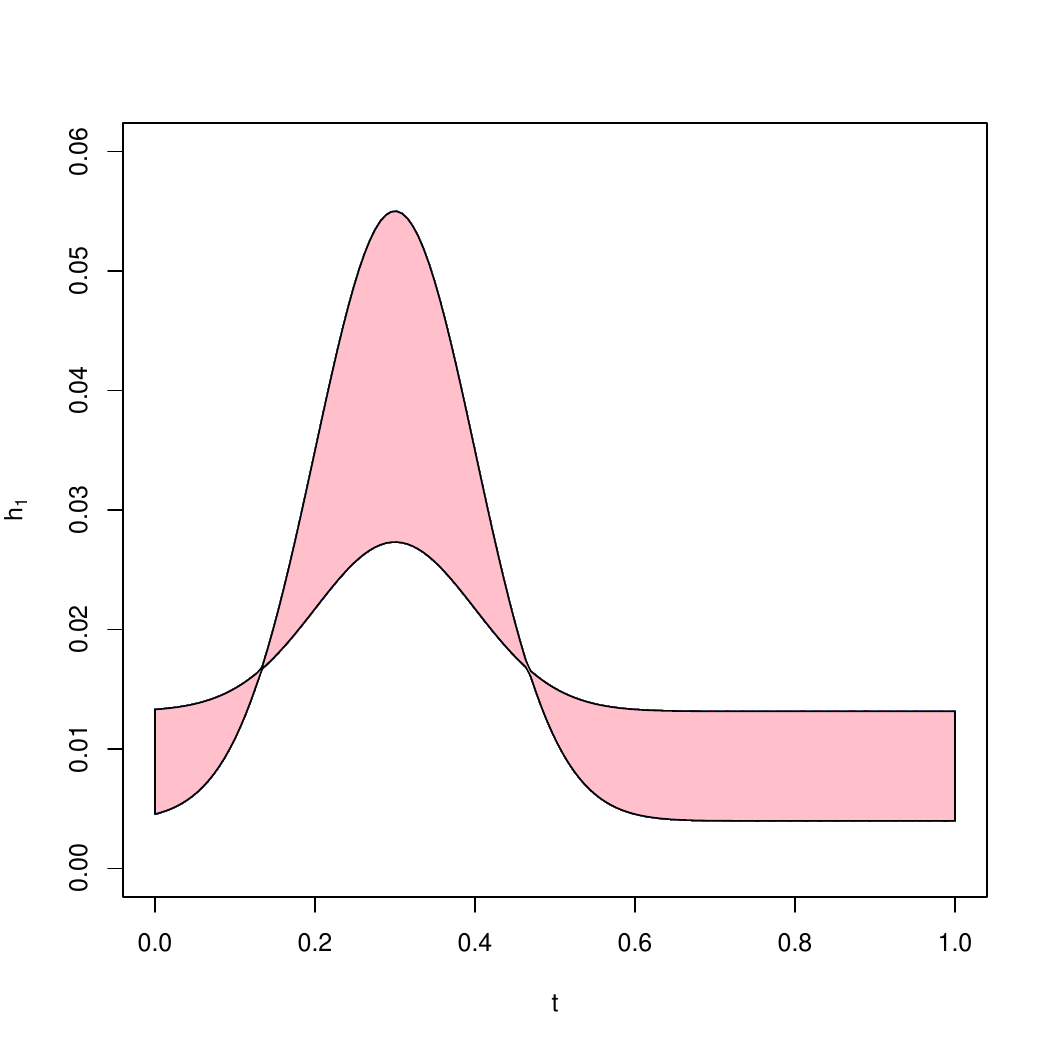} 
\end{tabular}
\end{center}
\caption{\em Two examples of finding bounds on the target signal density $h_1(t)$
in the sPlot model with no prior knowledge of $g_1$ and $g_2$.}
\label{fig::bounds}
\end{figure}

Two examples are shown in Figure \ref{fig::bounds}.
In the first example (left plot),
$g_1$ and $h_1$ are N(.5,.1)
while
$g_2$ and $h_2$ are Uniform(0,1).
In the second example (right plot),
$g_1$ is N(.7,1),
$g_2$ is Beta(3,2),
$h_1$ is N(.3,.1) and
$h_2$ is Uniform(0,1).
We see that the bounds on $h_1$ are tighter in the first example.
An interesting open question is to determine what
conditions cause the bounds to be narrow or wide.

\subsection{Optimal Weight Function}

\cite{dembinski2022} found the optimal weight function
for estimating $z$ in the sPlot model, but it is also
interesting to consider the estimation of $h_1(t)$.
Assume again the case $s=b=1$.
Let $f(t)$ be a fixed function and suppose we want to estimate
$\psi = \int f(t) h_1(t)dt.$
Note that $\psi = \E[w(M)f(T)]/\E[w(M)]$
so the estimate is
$$
\widehat\psi = 
\frac{\sum_i f(T_i)w(M_i)}{\sum_i w(M_i)} = \frac{\sum_i A_i}{\sum_i W_i},
$$
where
$A_i = f(T_i)w(M_i)$ and
$W_i = w(M_i)$.
Then
$\E[A] = z\psi$
and $\E[W]=z$ and
$$
\sqrt{n}(\hat\psi - \psi)\rightsquigarrow N(0,\sigma^2(w))
$$
where
$\sigma^2(w) = {\bf v}^T \boldSigma {\bf v}$,
${\bf v} = (1/z,-\psi/z)^T$ and
$$
\boldSigma =
\left(
\begin{array}{cc}
\mathrm{Var}[A]   & \mathrm{Cov}[A, W] \\
\mathrm{Cov}[A, W] & \mathrm{Var}[W]
\end{array}
\right) =
\left(
\begin{array}{cc}
\E[f^2(T)w^2(M)] - \psi^2 z^2 & \E[f(T)w^2(M)] - \psi z^2\\
\E[f(T)w^2(M)] - \psi z^2     & \E[w^2(M)] - z^2
\end{array}
\right).
$$
The functional derivative of $(1/2)\Sigma$ with respect to $w(m)$ is
$$
\boldSigma' =
w(m)
\left(
\begin{array}{cc}
p(m)\ell_2(m) & p(m)\ell_1(m)\\
p(m)\ell_1(m) & p(m)
\end{array}
\right) \equiv w(m) \boldLambda,
$$
where
$$
\ell_1(m) = \E[f(T)|M=m],\ 
\ell_2(m) = \E[f^2(T)|M=m].
$$
We find the stationary point of
$$
\frac{1}{2}\frac{\delta\sigma^2(w)}{\delta w}
-\alpha_1 \int w(m)g_1(m)dm - \alpha_2 \int w(m)g_2(m)dm,
$$
which leads to
${\bf v}^T \boldSigma' {\bf v} - \alpha_1 g_1(m) - \alpha_2 g_2(m) = 0$
and hence
$w(m) {\bf v}^T \boldLambda {\bf v}  - \alpha_1 g_1(m) - \alpha_2 g_2(m) = 0.$
The solution is
$$
w(m) = \frac{\alpha_1}{{\bf v}^T \boldLambda {\bf v}} g_1(m) + \frac{\alpha_2}{{\bf v}^T \boldLambda {\bf v}} g_2(m),
$$
where
$\boldalpha = {\bf Q}^{-1} {\bf e}_1$
with
$$
Q_{ij} = \int \frac{g_i(m)g_j(m)}{{\bf v}^T \boldLambda {\bf v}} \,dm.
$$

Now consider estimating $h_1(t_0)$.
We use the weighted kernel estimator
$$
\hat h_1(t_0) = \frac{1}{\sum_i w(M_i)} \sum_i K_\nu (T_i-t_0)w(M_i),
$$
where $\nu$ is the bandwidth.
The optimal $w$ is as above with
$f(t) = K_\nu(t-t_0)$.
However, this depends on the unknown $h_1(t)$.
One could then iterate between estimating $h_1$ and finding $w$.

\section{Discussion and Conclusion}

Separating background from signal
is an important task in particle physics.
The sPlot and COWs methods are
powerful and commonly used tools.
These methods allow
one to leverage information
on one variable $M$ to help
disentangle the signal and background
for another variable $T$.
We have provided the first 
investigation of these methods
from a formal statistical perspective.

In the absence of extra information on $M$,
the sPlot model is not identified
but it is still possible to bound the signal
distribution for $T$.
An interesting future project would be to 
more deeply understand these bounds and 
find conditions that imply the bounds are narrow.
For COWs, the situation is less clear and there are still
open questions about the identifiability of the model.
We have proposed an alternative approach
based on mixtures of copulas.
This allows us to relax the conditional independence assumption
while retaining identifiability as long as
$g_1$ and $g_2$ are estimable.
We also suggested a few other extensions including
a method for the case where the signal is localized.

In the case where there are three or more variables,
the model becomes nonparametrically identified.
We have indicated how the sPlot method
can be used here to provide a new algorithm
for estimating the model.
In future work, we will compare this approach to
existing algorithms.

\section*{Acknowledgments}
The authors thank Michael Schmelling for multiple useful discussions. Thanks to Sara Algeri for organizing the PhyStat Informal Review seminar that facilitated this collaboration and to members of the STAMPS Research Center at CMU for helpful feedback on this work. LW and MK were partially supported by NSF grant DMS-2310632.

\bibliography{main}

\end{document}